\pdfoutput=1
\documentclass[10pt,amsmath,amssymb,aps,prx,superscriptaddress,floatfix,longbibliography,twocolumn]{revtex4-2} 
\usepackage{silence}
\usepackage{graphicx} 
\usepackage{hyperref}   
\usepackage{amsmath}
\usepackage{xr}
\usepackage[bb=libus]{mathalpha}
\usepackage{xcolor}
\usepackage[xparse]{tcolorbox}
\usepackage[normalem]{ulem}
\usepackage{lineno}
\usepackage{braket}
\usepackage{ulem}
\definecolor{purple1}{HTML}{a156f0}

\definecolor{blue1}{HTML}{3f76b5}

\definecolor{green}{HTML}{36ba2b}

\newcounter{para}

\newcommand{\xCornell}{Department of Physics, Cornell University, Ithaca, NY, USA}
\newcommand{\xumass}{Department of Physics, University of Massachusetts, Amherst, MA, USA}
\newcommand{\unige}{Department of Theoretical Physics, University of Geneva, 24 quai Ernest-Ansermet, 1211 Gen\`eve, Switzerland}
\newcommand{\xEwha}{Department of Physics, Ewha Womans University, Seoul, South Korea}

\begin{document}

\title{Learning measurement-induced phase transitions using attention}
\author{Hyejin Kim}
\affiliation{\xCornell}
\author{Abhishek Kumar}
\affiliation{\xumass}
\author{Yiqing Zhou}
\affiliation{\xCornell}
\author{Yichen Xu}
\affiliation{\xCornell}
\author{Romain Vasseur}
\affiliation{\unige}
\author{Eun-Ah Kim}\thanks{Corresponding author: eun-ah.kim@cornell.edu}
\affiliation{\xCornell}
\affiliation{\xEwha}

\begin{abstract}
Measurement-induced phase transitions (MIPTs) epitomize new intellectual pursuits inspired by the advent of quantum hardware and the emergence of discrete and programmable circuit dynamics. 
Nevertheless, experimentally observing this transition is challenging, often requiring non-scalable protocols, such as post-selecting measurement trajectories or relying on classical simulations.
We introduce a scalable data-centric approach using Quantum Attention Networks (QuAN) to detect MIPTs without requiring post-selection or classical simulation. Applying QuAN to dynamics generated by Haar random unitaries and weak measurements, we first demonstrate that it can pinpoint MIPTs using their interpretation as ``learnability" transitions, where it becomes possible to distinguish two different initial states from the measurement record, locating a phase boundary consistent with exact results. Motivated by sample efficiency, we consider an alternative ``phase recognition" task--classifying weak- and strong-monitoring data generated from a single initial state. 
We find QuAN can provide an efficient and noise-tolerant upper bound on the MIPT based on measurement data alone by coupling Born-distribution-level (inter-trajectory) and dynamical (temporal) attention. In particular, our inspection of the inter-trajectory scores of the model trained with minimal sample size processing test data confirmed that QuAN paid special attention to the tail of the distribution of the Born probabilities at early times. This reassuring interpretation of QuAN's learning implies the phase-recognition approach can meaningfully signal MIPT in an experimentally accessible manner. Our results lay the groundwork for observing MIPT on near-term quantum hardware and highlight attention-based architectures as powerful tools for learning complex quantum dynamics.
\end{abstract}
\maketitle

\begin{figure*}[t]
    \centering
    \includegraphics[width=1\linewidth]{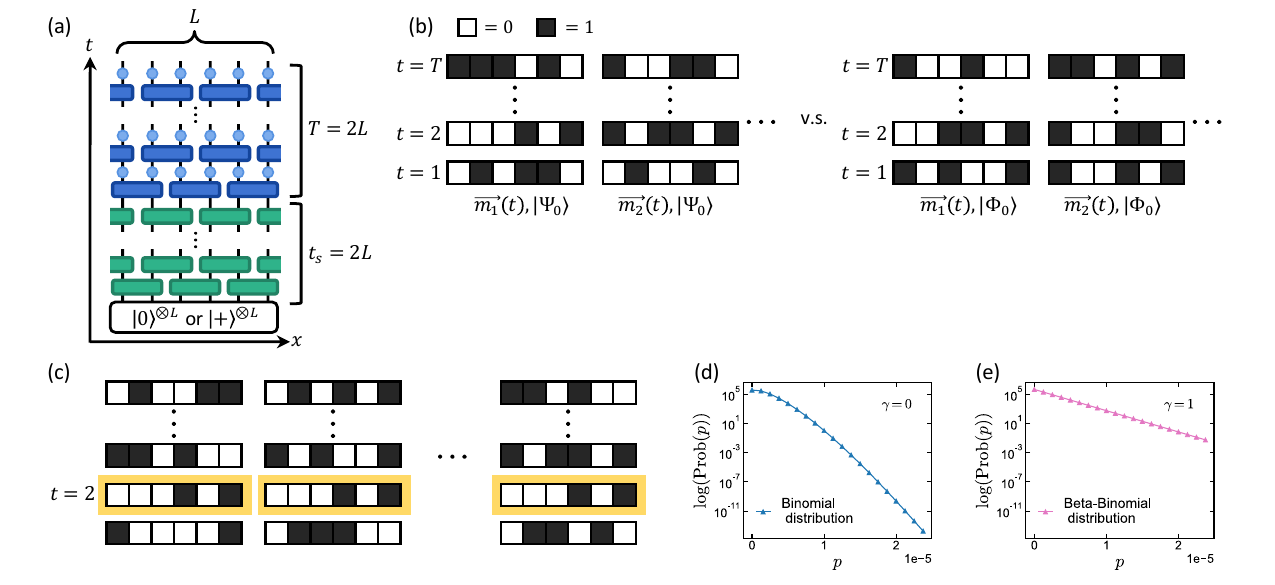}
    \caption{
{\bf The monitored quantum circuit and protocols for observing measurement-induced phase transitions (MIPTs) with weak measurements.} 
(a) An $L$-qubit quantum system prepared in the product state $|0\rangle^{\otimes L}$ or $|+\rangle^{\otimes L}$ (where $|+\rangle=\frac{1}{\sqrt{2}}(|0\rangle+|1\rangle)$) is scrambled through a brickwork circuit using a Haar unitary two-qubit gate $U_s$ (green block) for time $t_s=2L$, producing scrambled initial state $|\Psi_0\rangle$ or $|\Phi_0\rangle$. Then, a single ancilla is used and reset for weak measurement (blue circle) at each qubit between successive unitary layers, with a Haar unitary two-qubit gate $U_m$ (blue block) applied for a time $T=2L$. See SM section C for the two-qubit gates $U_s$ and $U_m$.
(b) A set of measurement trajectories $\vec{m}_1(t), \vec{m}_2(t), \cdots$ consists of bitstrings of length $L$ at each time $t$. 
(c) Repeated experiment yields a collection of trajectories $\{\vec{m}(t)\}_{|\psi\rangle}$ for the given initial state $|\psi\rangle$.  
(d,e) Contrasting distributions ${\rm Prob}(p)$ of Born probabilities estimated from a sample size $M$ for a system of size $L$ with zero net $\sigma_z$ expectation value of ancilla outcome. (d) For a system of unentangled qubits ${\rm Prob}(p)$ follows binomial distribution. (e) In the strong measurement limit, ${\rm Prob}(p)$ follows the Beta-binomial distribution with a longer tail.
}
    \label{fig:1}
\end{figure*}

Remarkable experimental developments with quantum hardware inspired the community to elevate the ``measurement problem'' ~\cite{Maudlin_1995} from a philosophical issue to a quantitative research frontier: the topic of monitored quantum dynamics. While general unitary dynamics scramble local quantum information, local measurements aim to extract information. The inherent tension in monitored dynamics that combine unitary evolution and intermittent measurements has been predicted to drive 
measurement-induced phase transitions (MIPTs)~\cite{li2018Phys.Rev.B, li2019Phys.Rev.B, skinner2019Phys.Rev.X, chan2019Phys.Rev.B, gullans2020Phys.Rev.X, zabalo2020Phys.Rev.B, bao2020Phys.Rev.B, jian2020Phys.Rev.B,PhysRevB.103.174309, potter2022entanglement, fisher2023random}. 
At a low measurement rate, the state evolved under monitored dynamics from a scrambled state (see Fig.~\ref{fig:1}(a)) is expected to remain volume-law entangled. However, measurements are expected to dominate the dynamics at a high measurement rate, purifying the volume-law state into an area-law-entangled state with reduced entanglement.

Although the field sprang from the inspirations of the experimental developments in quantum hardware, entanglement entropy scaling of the post-measurement state as the barometer of MIPT, as described above, faced the challenge of the exponential cost of post-selection. This challenge limited the experimental realizations of MIPT to classically simulatable small systems~\cite{noel2022Nat.Phys., hoke2023Nature, koh2023Nat.Phys.}.
One proposed strategy for avoiding post-selection was to compare quantum measurements against classical simulations~\cite{li2023Phys.Rev.Lett., garratt2024PRXQuantum,yanay2024, kamakari2024,feng2025postselection}, which restricts experimental study to circuits that are efficiently simulatable on classical computers. Alternatively, recently proposed  ``learnability'' perspective shifts the focus to information extracted from measurements~\cite{barratt2022Phys.Rev.Lett., dehghani2023NatCommun,ippoliti2024PRXQuantum,PhysRevB.109.094209,agrawal2024Phys.Rev.X, McGinley2024PRXQuantum,Hu2025}.
From this perspective, MIPT is signaled by whether sufficient information is extracted in measurement for a ``decoder''~\cite{choi2020Phys.Rev.Lett.,gullans2020Phys.Rev.Lett.} to infer some distinguishing property of the initial state. For instance, the decoder's task could be to determine whether the measurement record came from one initial state or another initial state (see Fig.~\ref{fig:1}(b)).
Now, the challenge becomes determining the sample complexity required to detect MIPT by observing the evolution of the decoder's capability to learn as the measurement strength is increased, which will depend on the decoding strategy. 
In particular, it is not known how to achieve a sample-efficient decoder that can address MIPT in the regime beyond classical simulation. 

Here we propose using the Quantum Attention Network (QuAN)~\cite{kim2024} to observe a MIPT under translationally invariant hybrid dynamics with Haar unitary two-qubit gates in a brickwall setting and weak measurements (see Fig.~\ref{fig:1}(a-c)).
The guiding insight is that, in this setting, strong early-time measurements, $\vec{m}(t)$ for small $t$, reveal information about the distribution of Born probabilities $p \equiv |\langle \psi_m(t)|\psi_m(t)\rangle|^2$, ${\rm Prob}(p)$. Here $|\psi_m(t)\rangle$ is the unnormalized quantum trajectory associated with the record $m$. Estimating $p$ from outcome frequencies yields a distribution ${\rm Prob}(p)$ that, for a scrambled pure state $|\psi\rangle$, follows the Beta–binomial form (see Fig.~\ref{fig:1}(d–e) and SM section A).
Access to the tail of this distribution will allow one to distinguish between two different scrambled states. We will leverage QuAN's ability to capture the complexity of a scrambled state using the attention mechanism~\cite{kim2024}.
First, we will establish that QuAN can learn the learnability transition marked by distinguishing measurement trajectories from two initial scrambled states in the strong measurement phase. 
We will interpret QuAN's learning through ablation studies to reveal the role of attention mechanisms. 
We will then frame a ``phase-recognition'' task for QuAN to contrast the weak-measurement phase from the strong-measurement phase. This latter task can establish a necessary condition for observing MIPT with noise tolerance and sample complexity that is experimentally accessible without the need for classical simulation. 

For our study, we adopt a translationally invariant hybrid dynamics with an eye towards experimental implementation. Specifically, taking a system of size $L$, we first use two-qubit Haar-random unitary $U_s$  in a brickwork arrangement of depth $2L$ (see Fig.~\ref{fig:1}(a) and SM section C) to scramble a simple product state, either $|0\rangle^{\otimes L}$ or $|+\rangle^{\otimes L}$ where $|+\rangle\equiv\frac{1}{\sqrt{2}}\left(|0\rangle+|1\rangle\right)$. 
The scrambled state, either $|\Psi_0\rangle$ or $|\Phi_0\rangle$, then goes through layers of entangling using a fixed Haar-random unitary $U_m$ followed by weak measurements of each qubit employing an ancilla~\cite{szyniszewski2019entanglement,bao2020Phys.Rev.B,agrawal2024Phys.Rev.X,aziz2024critical}. To perform a weak measurement of a system qubit $Q$ at time $t$, the system qubit is entangled to the ancilla qubit by a varying degree $\gamma$ using a two-qubit unitary $U_c(\gamma)$ and then the ancilla is projectively measured in the computational basis to get binary measurement outcomes $0$ or $1$. The ancilla is reset before repeating the process for the next qubit, until the weak measurement is completed for the entire system (see Fig.~\ref{fig:1}(b)). 

\begin{figure*}[t]
    \centering
    \includegraphics[ width=1\linewidth]{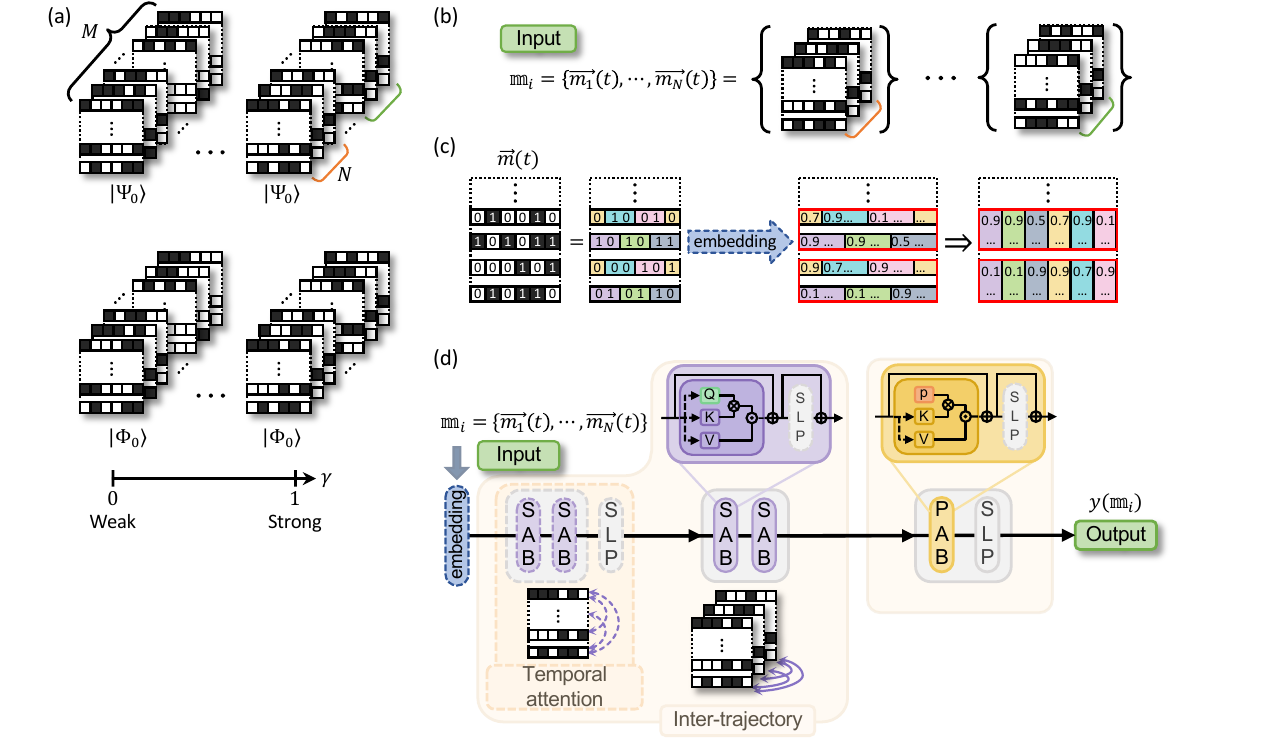}
    \caption{
    {\bf The data structure and the model architecture (QuAN) for detecting measurement-induced phase transitions from a set of trajectories.}
    (a) Data structure. For each initial state $|\Psi_0\rangle$ or $|\Phi_0\rangle$ and measurement strength $\gamma$, we prepare $M$ measurement trajectories ${\vec{m}(t)}$.
    (b) From the $M$ trajectories, we construct sets of size $N$, denoted by $\mathbb{m}_i = \{\vec{m}_1(t), \vec{m}_2(t), \cdots, \vec{m}_N(t)\}_i$.
    (c) The embedding layer encodes the spacetime position of each measurement, respecting the brickwork structure of the circuit, by pairing neighboring qubits within the same entangling gate. Each pair is mapped to a continuous vector, and two consecutive time steps ($t=2k-1$ and $t=2k$ for $k=1,\cdots,L$) are concatenated to ensure time-translational invariance.
    (d) Schematic of the architecture. After the embedding layer, each input set $\mathbb{m}_i$ passes through two layers of temporal and inter-trajectory attention, where each attention mechanism can inspect temporal correlations and inter-trajectory correlations.
    The decoder compresses set-structured data using the pulling attention block (PAB) module and a single-layer perception, in a set-permutation-invariant manner. The binary model output $y(\mathbb{m}_i)=0,1$ is associated with two different initial states or with weak- or strong-monitoring phases, respectively, for the initial state distinguishing or phase recognition tasks.
    }
    \label{fig:2}
\end{figure*}

Under this weak measurement setting, most of the information about the initial state, if any, is extracted at early times. In the weak monitoring limit, the ancilla is oblivious of the system at all times. Moreover, the ancilla does not have dynamics on its own.  
Hence, the distribution of the measurement outcome frequencies at different times (see Fig.~\ref{fig:1}(c)) will only reflect the initial state of the ancilla and exhibit no time dependence. In this case, if the ancilla is initialized in the $\sigma_z$ eigenstate, the measurement outcome will be highly concentrated. If the ancilla is initialized in a state with net zero magnetization, e.g., $\sigma_y$ eigenstate, the measurement outcome will be spread out, but the ancilla and the system are unentangled. Then the distribution ${\rm Prob}(p)$ of the probability $p$ will follow a binomial distribution $B(M,1/2^L)$, when the probability $p$ of a particular bitstring (measurement record) is estimated as the fraction of bitstring occurrence out of $M$ samples (see Fig.~\ref{fig:1}(d) and SM section A1).

In the strong monitoring limit, the projective measurements of the ancillas will also project the system state, completely extracting the knowledge of the initial state after the first round of measurements. When the ancillas efficiently extract information from the scrambled state of the system, the frequency of a particular bitstring within a sample size of $M$ will follow the Beta-binomial distribution (see Fig.~\ref{fig:1}(e) and SM section A2) with a longer tail, meaning that some small fraction of bitstrings occur with higher frequencies.
However, after early measurements, the projected system qubits will be short-range entangled, having only one layer of entangling gates $U_m$. Now ${\rm Prob}(p)$ will again follow a binomial distribution $B(M,1/2^L)$. Hence ${\rm Prob}(p)$ will evolve as a function of time from Beta-binomial distribution to binomial distribution in the strong monitoring limit. 
The mission for QuAN will be to learn enough about the evolving distributions ${\rm Prob}(p)$ to distinguish between the different initial states. 
To see whether such learning is possible, it is instructive to consider the average conditional probability for inferring the correct initial state $\psi^*$ given the measurements $P_{\rm corr}$ defined as~\cite{barratt2022Phys.Rev.Lett.,agrawal2024Phys.Rev.X}~\footnote{Ref.~\cite{agrawal2024Phys.Rev.X} referred to this quantity as {\it credence}.} 
\begin{equation}\label{eq:pcorr}
    P_{\rm corr} \equiv \mathbb{E}[p(\psi^*|m)] = \mathbb{E}\left[\frac{p(m|\psi^*)}{p(m|\Psi_0) + p(m|\Phi_0)}\right].
\end{equation}
Here $\mathbb{E}[\cdot]$ denotes the average over all measurement outcomes, and $\psi^*$ is the correct initial state label. As we show in SM Section B, $P_{\rm corr}=2/3$ with an infinitely large set of measurements in the thermodynamic limit, hence the state-distinguishing should be possible with a powerful enough decoder using a large enough volume of data.
The challenge will then be in accomplishing this decoding with a finite sample size.

In earlier efforts to detect MIPT using machine learning (ML), the architectures were designed to learn spatio-temporal correlations in the individual trajectories. In Ref.~\onlinecite{dehghani2023NatCommun}, the spacetime trajectories $\vec{m}(t)$ were treated as a two-dimensional image and fed into a convolutional neural network, which excels at learning local correlation to recognize images. 
\textcite{agrawal2024Phys.Rev.X} employed a recurrent neural network, well-suited for modeling temporal sequences, to capture the structure whereby the bitstring at time $t$ is conditioned on bitstrings from earlier times. However, when the trajectories are individually fed into an architecture, the ML model can only learn 
the distribution ${\rm Prob}(p)$ indirectly.
Consequently, the success of such efforts was limited to noiseless Clifford circuits~\cite{dehghani2023NatCommun} or charge-preserving dynamics~\cite{agrawal2024Phys.Rev.X}, which adds a global classical label (charge) to quantum states that is easier to learn. To go beyond these limiting cases and aim for an experimental signal of MIPT in the regime inaccessible to classical simulation, we employ QuAN~\cite{kim2024} with temporal attention and inter-trajectory attention. The inter-trajectory attention will offer an immediate comparison between different trajectories with high or low Born probability $p$, and temporal attention will enable learning the dynamical evolution of ${\rm Prob}(p)$.

QuAN was first introduced in Ref.~\onlinecite{kim2024} as an architecture for comparing measurement outcomes of two complex many-body states that cannot be distinguished by any quantity linear in their density matrices. 
Initially designed for measurements from a static density matrix, QuAN uses the self-attention mechanism that powers large language models~\cite{vaswani2017attention}, which learns the probability distribution of words by weighing the importance of correlations between words. The key insight was to input batches of final measurement outcomes as a set~\cite{lee2019set}, ensuring permutationally invariant treatment of bitstrings, while layers of self-attention between bitstrings in the set can learn moments of the bitstring distributions.

To train and test QuAN, we sample $M$-measurement trajectories $\{\vec{m}(t)\}$ starting from each initial state $\ket{\Psi_0}$ and $\ket{\Phi_0}$ at varying measurement strengths $\gamma$ (see Fig.~\ref{fig:2}(a)). We take care to organize and embed data to maximize the availability of meaningful information to QuAN. Firstly, we batch the entire sample into sets of size $N$ to afford direct access to the distribution properties of the trajectories, as in Ref.~\onlinecite{kim2024} (see Fig.~\ref{fig:2}(b)). To reflect the brickwork structure of the circuits, we pair up the measurements from qubits that have just undergone the entangling and concatenate two time steps to form a time-translationally invariant embedding (see Fig.~\ref{fig:2}(c)). Each embedded batch is fed into QuAN as a set of size $N$.
Each set $\mathbb{m}=\{\vec{m}_1(t),\cdots ,\vec{m}_N(t)\}$ goes through two layers of temporal self-attention and then two layers of inter-trajectory self-attention before the pooling layer and a single-layer perceptron collects all the learning into a scalar output $y(\mathbb{m})$ (see Fig.~\ref{fig:2}(d) and SM section E1 for details). The hyperparameters that specify the function $y(\mathbb{m})$ are determined through stochastic gradient descent that minimizes the binary cross entropy between the ground truth label $Y$ and the QuAN output $y(\mathbb{m})$ (see SM section E2 for details). 
To distinguish between initial states, $Y$ will be either $0$ or $1$ depending on the initial state.  
For phase recognition that distinguishes weak and strong monitoring phases associated with the same initial state, $Y$ will be $0$ for data from the weak monitoring phase and $1$ for the data from the strong monitoring phase. We denote the output of the trained model by $y^*(\mathbb{m})$ in both cases.

\begin{figure*}[t]
    \centering
    \includegraphics[width=\linewidth]{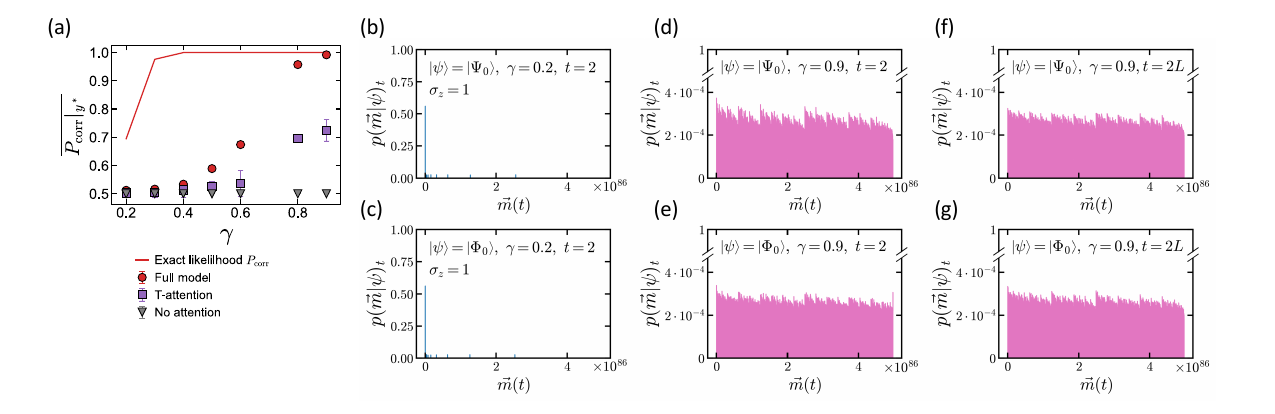}
    \caption{
    {\bf QuAN's learning of learnability transition. }
    (a) Average likelihood of inferring the correct initial state $\overline{P_{\rm corr}\vert_{y^*}}$ as a function of weak measurement strength $\gamma$. The plot compares performance across three architectures: an architecture that utilizes both temporal and intra-trajectory attention (red circles), an architecture with only temporal attention (purple squares), and an architecture without attention (gray triangles), where all three architectures use an optimal set size of $N=64$. For reference, we compare with the exact (optimal) likelihood $P_{\rm corr}$ derived from a noiseless monitored circuit simulation (red line). The analysis utilizes $M=6,000,000$ noiseless samples with a system size of $L=12$. The mean and errors are obtained from 4 independent model training runs.
    (b-g) Born probabilities of the measurement outcome from two initial states. 
    (b,c) Born probabilities of early-time $t=2$ measurement outcome at $\gamma=0.2$ for initial states  (b) $\ket{\Psi_0}$ and (c)  $\ket{\Phi_0}$. 
    (d,e) Born probabilities of early-time $t=2$ measurement outcome at $\gamma=0.9$ for initial states each (d) $\ket{\Psi_0}$ and (e)  $\ket{\Phi_0}$. 
    (f,g) Born probabilities of late-time $t=2L$ measurement outcome at $\gamma=0.9$ for initial states each (f) $\ket{\Psi_0}$ and (g)  $\ket{\Phi_0}$. }
    \label{fig:3}
\end{figure*}

For the state distinguishing task, we consider two product states $|0\rangle^{\otimes L}$ and $|+\rangle^{\otimes L}$, where $|+\rangle=\frac{1}{\sqrt{2}}\left(|0\rangle+|1\rangle\right)$. After the Haar-random scrambling evolution (green layers in Fig.~\ref{fig:1}(a)), there is little trace of the initial product state that can be gleaned from individual measurement outcomes spanning the enormous Hilbert space. The task at hand is to predict the correct initial state between $|\Psi_0\rangle$ and $|\Phi_0\rangle$, which respectively evolved from $|0\rangle^{\otimes L}$ or $|+\rangle^{\otimes L}$ under scrambling, given the measurement collection $\{\vec{m}(t)\}$ consisting of $M/N$ sets, $\mathbb{m}_i$. The trained model's average outcome $\langle{y^*}\rangle_{\{\vec{m}(t)\}}$ amounts to the model's prediction of the conditional probability of initial state $\Phi_0$, $P(\psi=\Phi_0|\{\vec{m}(t)\})$, i.e.,
\begin{equation}
    P(\psi=\Phi_0|\{\vec{m}(t)\})\approx\langle{y^*}\rangle_{\{\vec{m}(t)\}}\equiv \frac{N}{M}\sum_{i}y^*(\mathbb{m}_i),
\end{equation}
where the `trained' model $y^*$ is chosen by minimizing testing loss along the training epoch (see SM section E2 for more details).
We can also generalize the average likelihood $P_{\rm corr}$ (see Eq.~\eqref{eq:pcorr}) that the trained QuAN's output, $y^*$, infers the correct state for multiple sets as
\begin{equation}
    P_{\rm corr}|_{y^*}\equiv\frac{N}{2M}\sum_{i}\left[Y_iy^*(\mathbb{m}_i)+(1-Y_i)(1-y^*(\mathbb{m}_i))\right].
\end{equation}
To assess robustness across different training runs, we train 4 times for each $\gamma$ and report the training average $\overline{P_{\rm corr}|_{y^*}}$ and the standard deviation of the likelihood in Fig.~\ref{fig:3}(a).
Clearly, decoding will be impossible in the weak monitoring limit, since the ancilla never extracts any knowledge of the system. Even in the strong monitoring limit, this decoding is impossible with a small collection of measurement trajectories, since each measurement outcome individually will only reflect the scrambled nature of the system state. 
However, in the limit of infinitely large samples, the exact likelihood of a correct inference based on early measurement ($t=1$) in the limit of strong measurement $\gamma=1$ approaches $P_{\rm corr}\rightarrow 1$ with large enough set size, i.e., $N\gg 1$ (see SM section B).  

\begin{figure*}[t]
    \centering
    \includegraphics[width=\linewidth]{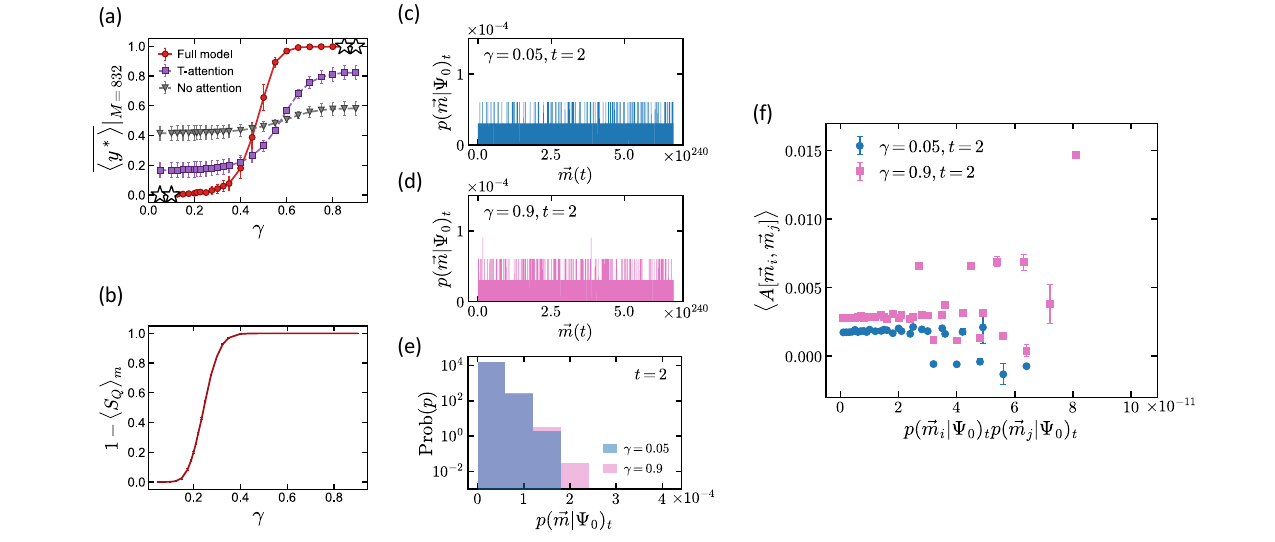}
    \caption{
    {\bf QuAN's learning of MIPT through phase recognition. }
    (a) Average strong monitoring phase prediction $\overline{\langle{y^*}\rangle}$ as a function of weak measurement strength $\gamma$. The plot compares performance across three architectures: one with both temporal and inter-trajectory attention (red circles), one with only temporal attention (purple squares), and one with no attention (gray triangles). We use $M=832$ trajectories per $\gamma$ from a noiseless monitored circuit with system size $L=20$, and optimal set size $N=64$ for demonstration. Star symbols indicate training points, and error bars represent the standard error of $\overline{\langle{y^*}\rangle}$ over 8 independently trained models.
    (b) Ancilla order parameter $1 - \langle S_Q \rangle_m$ as a reference for the transition. $S_Q$, the von Neumann entanglement entropy between the ancilla and the system, is computed by coupling an ancilla qubit to the initial state $\ket{\Phi_0}$ and applying scrambling followed by monitored dynamics.
    (c) Born probabilities of early-time ($t=2$) measurement outcomes at $\gamma=0.05$ for phase recognition setup with initial ancilla in $\sigma_y=1$ eigenstate. 
    (d) Born probabilities of early-time ($t=2$) measurement outcomes at $\gamma=0.9$.
    (e) Distribution of Born probability $\mathrm{Prob}(p)$, where $p \equiv p(\vec{m}|\Psi_0, \gamma)_{t=2}$ is the probability of a measurement outcome at early time $t=2$. $\mathrm{Prob}(p)$ follows a binomial distribution for $\gamma=0.05$ and a Beta-binomial distribution for $\gamma=0.9$ at early time.
    (f) The average inter-trajectory attention score $\langle A[\vec{m}_i,\vec{m}_j]\rangle$ plotted against the product of Born probabilities of the pair of measurement outcomes at early time $t=2$, $p(\vec{m}_i|\Psi_0)_tp(\vec{m}_j|\Psi_0)_t$, for weak (pink) and strong (blue) monitoring limits. The attention scores are evaluated using the QuAN trained with a minimal sample size of $M^*=832$ testing 50000 trajectories to obtain statistics.
 }
    \label{fig:4}
\end{figure*}

Fig.~\ref{fig:3}(a) presents QuAN's learning of the learnability transition for system size of $L=12$ with the ancilla initialized in the $\sigma_z=+1$ state. We use 7,000,000 trajectories for each initial state, split between training ($M=6,000,000$) and testing ($M=1,000,000$), and set size of $N=64$ (see SM section F1 for the set size optimization study and SM section F2 for the sample complexity study). 
We train QuAN for each value of $\gamma=0.2, 0.3, 0.4, 0.5, 0.6, 0.8, 0.9$ independently, and then estimate $P_{\rm corr}$ for that monitoring strength. 
The exact (optimal) likelihood of correct inference $P_{\rm corr}$ calculated by evaluating Born probabilities exactly (see SM section B) reaches 1 by measurement strength $\gamma_c=0.39$. 
QuAN's estimate $\overline{P_{\rm corr}|_{y^*}}$ starts to peel off from a random guess at $\gamma=\gamma_c$ to reach $\overline{P_{\rm corr}|_{y^*}}=1$ by $\gamma=1$. To gain insight into what was learned, we carry out ablation studies by removing different attention mechanisms. The results indicate that both temporal and inter-trajectory attention are essential.

To understand the success of QuAN, we compare the statistics of the measurement outcome from the two initial states at early and late times in Fig.~\ref{fig:3}(b-g). Specifically, we inspect the trajectories at a fixed time $t$ and count the frequency $k_{\vec{m}}$ of each bitstring outcome among $2^L$ possibilities of $\vec{m}$, for given initial state $|\psi\rangle=|\Psi_0\rangle,|\Phi_0\rangle$. Normalizing this frequency $k_{\vec{m}}$ with the total sample size $M$ yields the statistical estimation for $p(\vec{m}|\psi)_t$. 
At weak monitoring strength of $\gamma=0.2$, the ancilla measurements are largely oblivious to the system initial states at all times, only reflecting the ancilla initial state $\sigma_z=+1$. Hence the Born probability $p(\vec{m}|\psi)_t$ is concentrated entirely at $(1,\cdots,1)$ at all times. 
Fig.~\ref{fig:3}(b,c) shows that the measurement outcomes
$p(\vec{m}|\psi)_{t=2}$ are identical between the two system initial states and highly concentrated at the ancilla initial state even though $\gamma\neq0$. Indeed, the distribution is time-independent at small $\gamma$ as shown in SM section D1. 
With strong monitoring, the ancilla extracts information from the system state at early times; hence, the measurement reflects system dynamics.
Specifically, most of the information regarding the system's initial state $\psi$ is extracted through measurements at early times. Indeed, although both measurement distributions spread over the full Hilbert space of outcomes, the early-time Born probabilities $p(\vec{m}|\Psi_0){t=2}$ and $p(\vec{m}|\Phi_0){t=2}$ remain distinct, as different outcomes occur with higher relative frequencies (see Fig.~\ref{fig:3}(d,e)). The distinction disappears in the long-time limit (see Fig.~\ref{fig:3}(f,g)). Inter-trajectory attention exposes the shape of Born probabilities, and temporal attention allows QuAN to learn the time evolution and leverage early time contrasts.
Hence, QuAN achieves optimal learning even with large samples, only when both attention mechanisms are in place.

Despite QuAN's success above, learning MIPT through state distinguishing requires prohibitively high sample complexity (see SM section F2 for the sample complexity study). 
This is because the difference between the measurement outcomes of the scrambled state is only accessible in the strong monitoring phase at short times, and even then, only through a small set of bitstrings that occur with slightly higher frequency due to the tail in the Beta-binomial distributed ${\rm Prob}(p)$. The measurement outcomes associated with different initial states only differ in which specific bitstrings occur with higher frequency, as the comparison between Fig.~\ref{fig:3}(d) and Fig.~\ref{fig:3}(e) shows. 
However, it may be possible for QuAN to learn the contrast between the weak-monitoring outcome and the strong-monitoring outcome 
with much lower sample complexity by learning to contrast the shape of ${\rm Prob}(p)$, specifically the existence of the ``longer tail'' of ${\rm Prob}(p)$ for higher Born probability $p$.
Hence, we embark on this ``phase-recognition'' task with measurement trajectories based on the same initial state $|\Psi_0\rangle$.
To avoid QuAN learning to measure the $\sigma_z$ expectation value of the ancilla, we consider a slightly different weak-measurement protocol in which the $\gamma=0$ limit ancilla state is a $\sigma_y=1$ eigenstate. 
Even so, the brick-wall circuit dynamics of the system qubit will leave a fingerprint of spatiotemporal correlations in the measurement record to an increasing degree as the measurement strength $\gamma$ increases. For our particular setup with the periodic boundary condition, $\vec{m}_i(t)$ and $\vec{m}_i(t+L)$ will be correlated within a given trajectory, due to the light-cone touching the system boundary.
Such an inevitable, gradually increasing finite-size correlation is, in principle, not related to MIPT physics (see SM section C3 for more details); yet, it will be seen by QuAN. On the other hand, as QuAN's success with the learnability transition implies, QuAN is capable of learning the statistics of ${\rm Prob}(p)$, which separates out early time strong-monitoring phase data from the rest through longer tails.

To assess the feasibility of phase recognition beyond the classically simulatable regime, we consider the system size that pushes against the limits of classical simulatability, $L=20$. 
Now we train QuAN using trajectories for this system obtained from two limiting monitoring strengths with label $Y=0$ for data from the weak-monitoring limit ($\gamma=0.05$ and $0.1$), and $Y=1$ for data from the strong monitoring limit ($\gamma =0.85$ and $0.9$). Given the training points, we choose the `trained model' by minimizing the sum of testing losses evaluated at the four training points.
Once trained, the sample-averaged output of the trained model $\langle{y^*}\rangle$, given the input data from intermediate monitoring strength $\gamma$, is QuAN's confidence in the data belonging to the strong monitoring phase. 
Here, we train QuAN 8 times and obtain the training average $\overline{\langle y^*\rangle}$ and the standard deviation of the strong monitoring phase prediction as shown in Fig.~\ref{fig:4}(a). Indeed, we find that QuAN, with both inter-trajectory and temporal attention, can be trained to distinguish between the weak and strong monitoring phases with as few as $M=832$ measurement trajectories per training point. (See the minimal sample complexity study in SM section F3.) 
Fig.~\ref{fig:4}(a) shows this strong monitoring phase prediction at different monitoring strengths together with training points.
Once again, the ablation studies show that both inter-trajectory attention and temporal attention are critical for successful phase recognition.

\begin{figure*}[t]
    \centering
    \includegraphics[width=\linewidth]{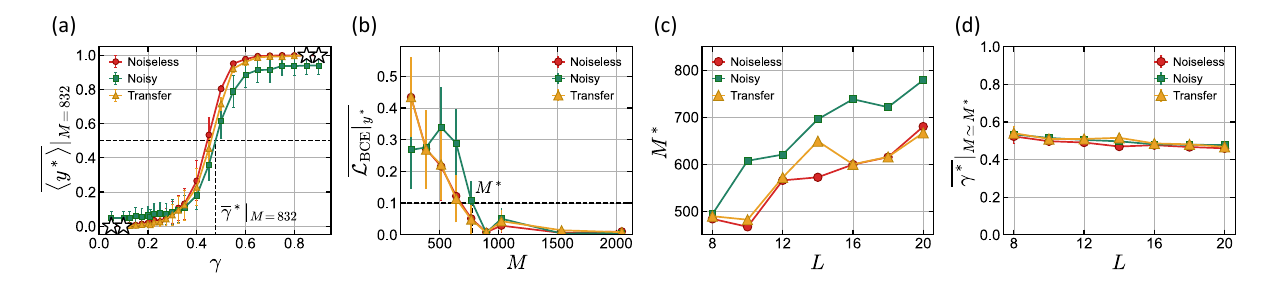}
    \caption{
    {\bf Impact of noise on MIPT phase recognition using QuAN.}
    (a) Average strong monitoring phase prediction $\overline{\langle y^*\rangle}$ for three training-testing setups using $M=832$ trajectories for training.  Star symbols mark training points, and error bars represent the standard deviation in strong monitoring phase prediction across 8 independent training runs. The horizontal dashed line $\langle y(\gamma^*)\rangle = 0.5$ is used to determine the location of the QuAN-estimated transition point $\gamma^*$.
    (b) Training averaged test loss $\overline{\mathcal{L}_{\rm BCE}|_{y^*}}$ as a function of training sample size $M$ for the three training-testing settings at $L=20$. We used the threshold value of $\overline{\mathcal{L}_{\rm BCE}|_{y^*}}=0.1$ to estimate the minimal sample complexity $M^*$ (see SM section F3 for details).
    (c) Minimal sample complexity $M^*$ as a function of system size $L$ for the three modules. Sample complexity grows faster for training on noisy data (blue) compared to training on noiseless data (red and purple). However, all three modules exhibit linear scaling of sample complexity with $L$.
    (d) Predicted transition point $\overline{\gamma^*}$ at minimal sample complexity $M \simeq M^*(L)$, plotted as a function of system size $L$.
  }
    \label{fig:5}
\end{figure*}

We can benchmark the phase recognition performance by QuAN with full attention against the trajectory-averaged von Neumann entropy of a reference qubit $\langle S_Q\rangle_m$, which was introduced in Ref.~\cite{gullans2020Phys.Rev.Lett.} as an effective order parameter for MIPT with the limiting value of $\langle S_Q\rangle_m=1$ in the weak monitoring phase and $\langle S_Q\rangle_m=0$ in the strong monitoring phase. Comparing QuAN's phase recognition (see Fig.~\ref{fig:4}(a)) upon training with just $M=832$ trajectories against the evolution of $1-\langle S_Q\rangle_m$ obtained using the full knowledge of the states and measurement outcomes (see Fig.~\ref{fig:4}(b) and SM section C2 for more details), we see that QuAN requires larger $\gamma$ to recognize strong monitoring phase. Finite-size scaling analysis of $\langle S_Q\rangle_m$ places the exact critical monitoring strength at $\gamma_{c} = 0.22(1)$.
Hence, the QuAN-based phase recognition will offer an upper bound on the critical measurement strength, i.e., $\gamma_c\leq \overline{\gamma^*}$, where $\gamma^*$ is the monitoring strength at which the average of the strong monitoring phase prediction $\langle{y^*}\rangle$ rises above 0.5, and obtain $\overline{\gamma^*}$ by averaging over 8 independent trainings.

The key question is whether QuAN extracted meaningful information from such a limited training sample for the phase recognition task. 
With the use of $\sigma_y=+1$ ancilla initial state in the weak monitoring limit, the measurement outcome is not concentrated even in the weak monitoring limit. Instead, the weak monitoring limit measurement outcome will sample between 0 and 1 with $50\%$ chance for each qubit, more or less independently. 
Hence ${\rm Prob}(p)$ for this short-range entangled state will follow the binomial distribution at all times (see Fig.~\ref{fig:4}(c) and Fig.~\ref{fig:4}(e)). On the other hand, the strong monitoring limit will reveal the volume-law entangled, scrambled system initial state through the longer tail of Beta-binomial distributed ${\rm Prob}(p)$ at early time (see Fig.~\ref{fig:4}(d) and Fig.~\ref{fig:4}(e)). As the system state becomes short-range entangled due to repeated nearly projective measurements at a late time, the strong monitoring regime data will also follow a binomial distribution.
Hence, to meaningfully contrast the weak and strong monitoring phases, it would be necessary to learn to recognize information from the longer tail of the scrambled state, which is only accessible at early times (see Fig.~\ref{fig:1}(e)).

To inspect what the model trained on $M=832$ trajectories per $\gamma$ learned, we study the inter-trajectory attention score associated $A[\vec{m}_i(t),\vec{m}_j(t)]$ with a pair of testing measurement outcomes $(\vec{m}_i(t),\vec{m}_j(t))$ at equal time $t$, indicated by double-headed arrow towards the bottom of Fig.~\ref{fig:2}(d). (see SM section E and G for technical definition.) 
The attention score depends on the learned hyperparameters of QuAN and the input data. To investigate how the learned hyperparameters discern the input data, we test the attention score using $M=50000$ trajectories and study how the attention score depends on the Born probability of each outcome. 
As shown in SM section G, 99.9\% of the inter-trajectory attention score $A[\vec{m}_i,\vec{m}_j]$ is distributed between -0.015 and 0.03. 
Averaging the attention score over a collection of measurement pairs $(\vec{m}_i, \vec{m}_j)$ with the same value for the product of their Born probabilities, $p(\vec{m}_i|\Psi_0)_tp(\vec{m}_j|\Psi_0)_t$ allows us to study how the ${\rm Prob}(p)$ influenced QuAN's learning. 
Such average $\langle A[\vec{m}_i,\vec{m}_j]\rangle$ plotted against $p(\vec{m}_i|\Psi_0)_t p(\vec{m}_j|\Psi_0)_t$ shown in Fig.~\ref{fig:4}(f) reveals that QuAN pays equal and low attention to outcomes with low Born probabilities, i.e., small $p(\vec{m}_i|\Psi_0)_t p(\vec{m}_j|\Psi_0)_t$. The outcomes that gain outsized attention are those with high Born probabilities for both outcomes, which only occur in the strong monitoring phase.  
This investigation provides reassuring confidence that the phase recognition learning that can be achieved with realistic sample complexity will detect MIPT in a meaningful manner.

To explore the feasibility of the above MIPT phase recognition on current quantum hardware, we study the impact of a realistic level of noise. Specifically, we consider a depolarizing noise model of the single-qubit and two-qubit gate error probabilities set at $p_{1q} \approx 4 \times 10^{-5}$ and $p_{2q} \approx 2 \times 10^{-3}$, respectively, in accordance with existing experimental error rates~\cite{pino2021demonstration}.
We now compare three training-testing setups: noiseless (using noiseless data for both training and testing), noisy (using noisy data for both training and testing), and transfer (training on noiseless data and testing on noisy data). The average strong monitoring phase prediction $\overline{\langle y^*\rangle}$ using QuAN with full attention trained with $M=832$ trajectories, shown in Fig.~\ref{fig:5}(a), presents an optimistic outlook. The fact that transfer learning prediction is almost identical to the noiseless learning prediction implies that experimental data can be tested using QuAN trained with simulated noiseless data. Most importantly, the fact that a noisy learning setting shows just as robust a phase recognition performance as the noiseless setting invites aiming for system sizes beyond classical simulations.

The feasibility of reaching system sizes beyond classical simulation will hinge on how the sample complexity scales. To estimate the sample complexity for MIPT phase recognition at larger system sizes, we define minimal sample complexity $M^*$ and study its scaling. For this, we train QuAN with varying sample sizes and chart the evolution of the test (binary cross-entropy) loss $\mathcal{L}_{\rm BCE}|_{y^*}$ for the trained model as a function of the sample size, for each system size. To capture variation over training runs, we average the test loss over 8 training runs to obtain the average test loss at the given sample size $\overline{\mathcal{L}_{\rm BCE}|_{y^*}}$. We then find the minimal sample complexity $M^*$ to be the sample size at which $\overline{\mathcal{L}_{\rm BCE}|_{y^*}}$ reaches below the threshold we chose to be 0.1. (See Fig.~\ref{fig:5}(b) for $L=20$ as an example, and SM section F3 for more details). 
As we show in Fig.~\ref{fig:5}(c), the resulting minimal sample complexity scales linearly as a function of sample size for all three training settings. Learning the tail of ${\rm Prob}(p)$ requires some degree of repetition and will ultimately require exponential cost in training. Nevertheless, QuAN learns meaningful distinctions with a sample size that scales linearly for the system sizes studied, suggesting the potential to surpass classical computation before exponential costs become prohibitive. Although a noisy training-testing setting has a slightly higher slope compared to the two settings trained with noiseless data, the linear scaling projects the sample complexity for system sizes beyond classical simulation at around $M\sim 1000$, with realistic noise, which is within experimental reach.
Compared to the exact threshold $\gamma_c=0.22$ calculated with the full knowledge of the system (see Fig.~\ref{fig:4}(b)), QuAN's critical monitoring strength at the minimal sample size, $\gamma^*|_{M=M^*}$, overestimates. Nevertheless, this estimate is consistently coming down with the increase in system size. Therefore, a phase recognition study with a minimal sample size can signal the existence of two phases and provide an upper bound for the threshold monitoring strength.

To summarize, we leveraged attention mechanisms through recently introduced QuAN~\cite{kim2024} to learn the MIPT without post-selection in a data-centric manner. The goal was to find a strategy that can be extended to system sizes beyond those that can be simulated classically. 
The weak-measurement setting we employed is natural for the learnability transition perspective of MIPT. The learnability transition occurs at the threshold monitoring strength required for the ancilla to be extracting enough information to distinguish between trajectories originating from two different initial states correctly. QuAN's success can be attributed to the two attention mechanisms hitting two critical aspects of the data problem. Firstly, the information ancilla has to extract is in the form of the distribution of measurement outcomes' frequencies. The inter-trajectory attention captures the shape of the distribution. Secondly, the amount of meaningful information extracted by the ancilla decays with time. The temporal attention tunes to this time-dependence. Building on this observation, we explored the phase-recognition perspective by training the same QnAN wth contrasting data sets from weak and strong monitoring. The phase-recognition required orders of magnitude lower sample complexity, and it proved to be robust against a realistic level of noise. Given the linear sample complexity growth we observed with a reasonable prefactor for the phase-recognition with QuAN, an exciting next step will be to probe MIPT with experimental data in the regime beyond classical simulation.

{\bf Code availability}
The code supporting the findings of this study is available at \url{https://github.com/KimGroup/MIPT_QuAN}.

{\bf Acknowledgements.}
We thank Henrik Dreyer, Sarang Gopalakrishnan, and  Andrew Potter for helpful discussions.
HK, YX, and E-AK acknowledge support from the NSF through OAC-2118310 and from the Gordon and Betty Moore Foundation’s EPiQS Initiative, Grant GBMF10436 to  E-AK. YZ acknowledges support from Platform for the Accelerated Realization, Analysis, and Discovery of Interface Materials (PARADIM), supported by the NSF under Cooperative Agreement No.\ DMR-2039380. AK and RV acknowledge partial support from the US Department of Energy, Office of Science, Basic Energy Sciences, under award No. DE-SC0023999.

\clearpage
\bibliography{combined}
\end{document}


\title{Supplementary materials for ``Learning measurement-induced phase transitions using attention''}
\author{Hyejin Kim}
\affiliation{\xCornell}
\author{Abhishek Kumar}
\affiliation{\xumass}
\author{Yiqing Zhou}
\affiliation{\xCornell}
\author{Yichen Xu}
\affiliation{\xCornell}
\author{Romain Vasseur}
\affiliation{\unige}
\author{Eun-Ah Kim}
\affiliation{\xCornell}
\affiliation{\xEwha}
\maketitle
\tableofcontents
\appendix

\section{Theoretical \texorpdfstring{$\mathrm{Prob}(p)$}{Prob(p)} for finite sample size \texorpdfstring{$M$}{M}}\label{sec:A}

In this section, we discuss the distribution of Born probabilities $ p(\vec{m}|\psi) = |\langle \vec{m}|\psi\rangle|^2$ for a scrambled pure state $|\psi\rangle$ at early time ($t=1$), $\Prob(p)\equiv\frac{1}{D}\sum_{\vec{m}}\delta(p(\vec{m}|\psi)-p)$, and compare the distribution between weak and strong measurement regimes.

\subsection{\texorpdfstring{$\Prob(p)$}{Prob(p)} at \texorpdfstring{$\gamma=0$}{γ=0} with ancilla prepared in \texorpdfstring{$\sigma_y=1$}{σ
y=1} eigenstate}\label{sec:A1}

First, we focus on the behavior of the distribution in the weak monitoring limit of $\gamma=0$. We consider the setting with ancilla initial state $\sigma_y=1$ eigenstate, as in the phase recognition task. See SFig.~\ref{fig:hybridsetup}(b) for the circuit setup. 
Here, projective measurements on each ancilla yield 0 or 1 in the computational basis with equal probabilities. 
Hence, all trajectories occur with equal probability $p = 1/D$, where $D = 2^L$ is the Hilbert space size for $L$ system qubits. The corresponding Born probability distribution is sharply peaked at $p = 1/D$:
\begin{equation}\label{eq:delta}
    \Prob(p) = \delta\left(p - \frac{1}{D}\right).
\end{equation}

Now consider estimating the distribution with a finite sample of size $M$. 
The count of a specific outcome, $k$, follows a binomial distribution, $k \sim B(M, 1/D)$.  Hence, the sampled estimate of the distribution is 
\begin{align}\label{eq: binom}
    \Prob\left(p = \frac{k}{M}; M\right) = M \begin{pmatrix} M \\ k \end{pmatrix} \left(\frac{1}{D}\right)^k \left(1 - \frac{1}{D}\right)^{M - k}.
\end{align}
We can verify that $\Prob(p)$ is properly normalized:
$\sum_{k=0}^M \Prob\left(\frac{k}{M}; M\right) \Delta p = \sum_{k=0}^M \Prob\left(\frac{k}{M}; M\right) \cdot \frac{1}{M} = 1$ \footnote{Note that for initial ancilla in $\sigma_z=1$ eigenstate which is for the state distinguishing circuit setup in SFig.~\ref{fig:hybridsetup}(a), $\Prob(p)$ follows delta distribution $\Prob(p) = \frac{D-1}{D}\delta(p) + \frac{1}{D}\delta(p-1)$.}. In the limit $M\to\infty $, the binomial distribution becomes the delta distribution in Eq.~\eqref{eq:delta}

\subsection{\texorpdfstring{$\Prob(p)$}{Prob(p)} at \texorpdfstring{$\gamma=1$}{γ=1}}\label{sec:A2}

In the strong monitoring limit of $\gamma=1$, the protocol projectively measures system qubits. For a random state of system size $L$ generated by scrambling dynamics, the Born probabilities follow the Beta distribution~\cite{Boixo2018NaturePhys}, given by
\begin{align}\label{eq:PTdist}
    \Prob(p) = (D - 1)(1 - p)^{D - 2} \to 2^L e^{-Dp} \quad \text{as } D \to \infty,
\end{align}
where $D = 2^L$ is given by the Hilbert space size.
The Beta distribution approaches an exponential distribution in the large-$D$ limit, commonly known as the Porter–Thomas distribution.

Now, we again consider 
estimating the distribution $\Prob(p)$ from a finite sample of size $M$.
$k$, the count of a specific measurement outcome follows $k \sim B(M, p)$, where $p \sim \Prob(p)$ is distributed according to the Beta distribution.
\begin{align}\label{eq:BB}
    \Prob\left(p = \frac{k}{M}; M\right) &= M \int_0^1 \binom{M}{k} p^k (1 - p)^{M - k} (D - 1)(1 - p)^{D - 2} dp \\
    &= M(D - 1)\frac{M! (M + D - k - 2)!}{(M - k)! (M + D - 1)!}.
\end{align}
This distribution is called the Beta-binomial distribution, where the parameter $p$ in the binomial distribution is randomly drawn from a beta distribution.
The normalization condition can be easily verified: $\sum_{k=0}^M \Prob(k/M;M)\Delta p = \sum_{k=0}^M \Prob(k/M;M)1/M = 1$.
In the large $M\to\infty$ limit, the Beta-binomial distribution returns to the Beta distribution,
\begin{align}
    \lim_{M\to\infty}\Prob\left(\frac{k}{M}; M\right) = (D - 1)(1 - p)^{D - 2},
\end{align}
to leading order, in Stirling's approximation. 

\subsection{Comparison between the tails of \texorpdfstring{$\Prob(p;M)$}{Prob(p)} at \texorpdfstring{$\gamma=0$}{γ=0} and \texorpdfstring{$\gamma=1$}{γ=1}}\label{sec:A3}

We aim to compare the tails of two distributions -- between $\gamma=0$ for ancilla prepared in $\sigma_y=1$ eigenstate and $\gamma=1$ -- at a finite sample size $M\ll D=2^L$. For each case, Born probability is given by Eq.~\eqref{eq: binom} (binomial, denoting as $\Prob_0(p)$) for $\gamma=0$, and Eq.~\eqref{eq:BB} (Beta-binomial, denoting as $\Prob_1(p)$) for $\gamma=1$. We assume $1/M\ll p\ll 1$ for a tail, and apply Stirling's approximation ($\ln N! = N \ln N - N + 1/2\ln(2\pi N)$):
\begin{align}
    \ln \Prob_0(p) &= \ln M-Mp\ln(Dp)-M(1-p)\ln(1-p) - M(1-p)\ln\left(\frac{D}{D-1}\right) - \frac{1}{2}\ln(2\pi Mp(1-p))+\ln M, \\
    \ln \Prob_1(p) &= \ln M + \ln(D-1) - M(1-p)\ln(1-p) + (M(1-p)+D-2)\ln\left(1+\frac{D}{M}-p-\frac{2}{M}\right) \\
    &\quad - (M+D-1)\ln\left(1+\frac{D}{M}-\cancel{\frac{1}{M}}\right) + \frac{1}{2}\cancel{\ln\left(\frac{M-Mp+D-2}{(1-p)(M+D-1)}\right)} - 1,
\end{align}
where we use slashes to denote terms that vanish in the limit $D\gg M \gg 1$. The logarithmic difference between two distributions is given by
\begin{align}
    \Delta(p) = \ln\Prob_0(p)-\ln\Prob_1(p) &= Mp\ln\left(\frac{M+D}{MDp}\right) + \ln\left(\frac{M+D}{MDp}\right) - \frac{1}{2}\ln\left(\frac{2\pi(1-p)}{Mp}\right)+1\\
    &\quad - (M(1-p)+D-2)\ln\left(1-\cancel{\frac{Mp+2}{M+D}}\right) - (M(1-p)-1)\cancel{\ln\left(\frac{D}{D-1}\right)} \\
    &= (1+Mp)\ln\left(1+\frac{M}{D}\right) - (1+Mp)\ln(Mp) - \frac{1}{2}\ln\left(\frac{2\pi(1-p)}{Mp}\right)+1.
\end{align}
Therefore, the ratio between two distributions at their tails is
\begin{align}
    \frac{\Prob_0(p)}{\Prob_1(p)} = e^{\Delta(p)} = \sqrt{\frac{e^2}{2\pi(1-p)}}\frac{(1+M/D)^{1+Mp}}{(Mp)^{1/2+Mp}}.
\end{align}
The above expression decreases as we increase $p$, eventually toward 0, implying $\Prob_1(p)$ has longer tail than $\Prob_0(p)$. The tail of $\Prob_0(p)$ is smaller than $\Prob_1(p)$ with the factor of $(1+M/D)^{1+Mp}\simeq 1+M(1+Mp)/D$ using Binomial approximation. This implies that for large $D=2^L$, $e^{\Delta}$ would approach 0 faster, resulting in a large difference between the tails of the binomial and Beta-binomial distribution.

\section{Optimal \texorpdfstring{$\mathcal{P}_\text{corr}$}{Pcorr} in the thermodynamic limit}
\label{sec:B}

Consider $P_\text{corr}$, the average likelihood of correctly inferring the initial state $\ket{\Psi_0}$ or $\ket{\Phi_0}$ based on an infinitely large volume of measurements.
We assume the observer can analytically compute Born probabilities (corresponding to a case of ``optimal decoding''), and provides a theoretical upper bound for how well any classifier can distinguish initial states. 

Suppose we perform a measurement and obtain outcome $m$. We determine the initial state using Bayes' theorem, where the posterior probability of being in state $\ket{\Psi_0}$ given outcome $m$ is
\begin{align}
    p(\Psi_0|m) = \frac{p(m|\Psi_0)}{p(m|\Psi_0)+p(m|\Phi_0)},
\end{align}
assuming equal priors, i.e., $p(\Psi_0) = p(\Phi_0) = 1/2$. The probability of inferring a correct initial state $\psi^*$ is then given by averaging the posterior probability over all possible outcomes and states~\cite{barratt2022Phys.Rev.Lett., agrawal2024Phys.Rev.X},
\begin{align}
    P_\text{corr} &= \mathbb{E}[p(\psi^*|m)] = \sum_{\{m\}} p(m|\Psi_0)p(\Psi_0)p(\Psi_0|m) + p(m|\Phi_0)p(\Phi_0)p(\Phi_0|m) \\
    &= \sum_{\{m\}} \frac{p(m|\Psi_0)^2}{p(m|\Psi_0)+p(m|\Phi_0)} = D \left\{ \frac{1}{D} \sum_{m=1}^{D} \frac{p(m|\Psi_0)^2}{p(m|\Psi_0)+p(m|\Phi_0)} \right\},
\end{align}
where $D= 2^L$. 
In the thermodynamic limit $D \to \infty$, 
\begin{align}
    \lim_{D\to\infty} P_\text{corr} = D \int_0^1 \int_0^1 \Prob(x)\Prob(y) \frac{x^2}{x+y} \, dx \, dy,
\end{align}
where we set $x \equiv p(m|\Psi_0)$ and $y \equiv p(m|\Phi_0)$, and $\Prob(x)$ denotes the distribution of Born probabilities. 
For $\gamma = 0$, where $\Prob(p) = \delta(p - 1/D)$, we can easily verify that $P_{\rm corr} = 1/2$. For $\gamma = 1$, where $\Prob(p)$ is given by Eq.~\eqref{eq:PTdist}, we find
\begin{align}
    \lim_{D\to\infty} P_{\rm corr} = \lim_{D\to\infty} D^3 \int_0^1 \int_0^1 e^{-D(x+y)} \frac{x^2}{x+y} \, dx \, dy = \frac{2}{3}.
\end{align}

A discretized quantity that defines success as $p(\psi^*|m)>1/2$ is an alternative metric referred to as 
the accuracy $\alpha$~\cite{barratt2022Phys.Rev.Lett., agrawal2024Phys.Rev.X}
\begin{align}
    \alpha &\equiv \mathbb{E}[p(\psi^*|m)>1/2],
\end{align}
where $\psi^*$ is again the correct initial state and $\mathbb{E}[\cdot]$ denotes the average over all measurement outcomes.
For $\gamma=1$ where $\Prob(p)$ is given by Eq.~\eqref{eq:PTdist}, we obtain in the thermodynamic limit, $\lim_{D\to\infty} \alpha = 3/4$. 
The two quantities -- $P_{\rm corr}$ and $\alpha$ -- are related but distinct, converging to different values at $\gamma=1$ in the thermodynamic limit. The main difference is that $P_{\rm corr}$ incorporates the magnitude of the posterior probability $p(\psi^*|m)$, whereas accuracy $\alpha$ records only whether the posterior exceeds 1/2 in a binary classification.

\begin{figure}[ht]
    \centering
    \includegraphics[page=1, width=0.9\linewidth]{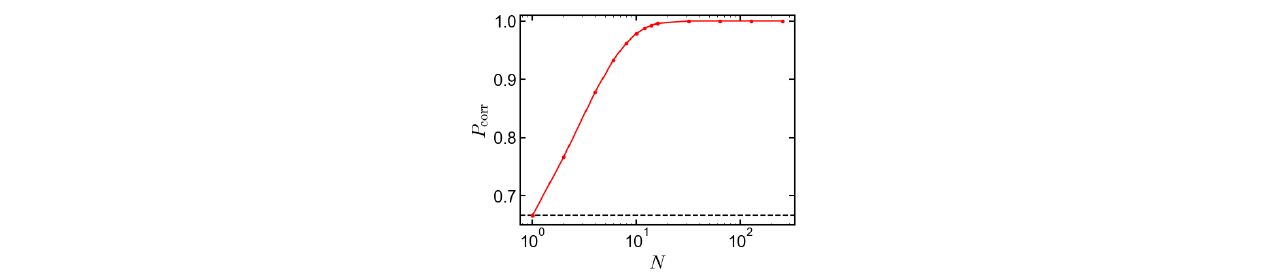}
    \caption{(a) Average optimal likelihood of correct initial state $P_{\rm corr}$ at $\gamma=1$, as a function of set size $N$. The optimal value of $P_{\rm corr}$ is calculated numerically using Monte Carlo integration (Eq.~\eqref{eq:pcorr_MC}). Dashed line represents $P_{\rm corr}=2/3$ using $N=1$. }
    \label{fig:1}
\end{figure}

To benchmark state-distinguishing performance by QuAN, which takes a measurement set as an input, consider the posterior probability to infer state $\Phi_0$ given the set of measurements $\{m\}= \{m_1, m_2, \cdots, m_N\}$: 
\begin{align}\label{eq:pcorr_N}
    p(\Psi_0|\{m\}) = \frac{p(\{m\}|\Psi_0)}{p(\{m\}|\Psi_0)+p(\{m\}|\Phi_0)} = \frac{\prod_{i=1}^N p(m_i|\Psi_0)}{\prod_{i=1}^N p(m_i|\Psi_0) + \prod_{i=1}^N p(m_i|\Phi_0)}.
\end{align}
Since measurements are independent and identically distributed, we have $p(\{m\}|\Psi_0) = \prod_{i=1}^N p(m_i|\Psi_0)$. The corresponding $P_{\rm corr}$ is
\begin{align}\label{eq:pcorr}
    P_\text{corr} = D^N \int_{[0,1]^N} \prod_{i=1}^N dx_i \int_{[0,1]^N} \prod_{i=1}^N dy_i \prod_{i=1}^N \Prob(x_i) \prod_{i=1}^N \Prob(y_i) \left( \frac{\left( \prod_{i=1}^N x_i \right)^2}{\prod_{i=1}^N x_i + \prod_{i=1}^N y_i} \right),
\end{align}
where we set $x_i \equiv p(m_i|\Psi_0)$ and $y_i \equiv p(m_i|\Phi_0)$. This integral is $2N$-dimensional and cannot be evaluated analytically. Instead, we estimate it using Monte Carlo integration with $M_{\rm mc}$ number of Monte Carlo samples.
\begin{align}\label{eq:pcorr_MC}
    P_\text{corr} = \lim_{M_{\rm mc}\to\infty} \frac{D^N}{M_{\rm mc}} \sum_{j=1}^{M_{\rm mc}} \left( \frac{\left( \prod_{i=1}^N x_i^{(j)} \right)^2}{\prod_{i=1}^N x_i^{(j)} + \prod_{i=1}^N y_i^{(j)}} \right).
\end{align}
SFig.~\ref{fig:1} shows $P_{\rm corr}$ for $1 \leq N \leq 256$, calculated via Monte Carlo integration using $M_{\rm mc} = 10^6$ samples. We observe that $P_{\rm corr} = 2/3$ for $N = 1$, and it further increases as the set size $N$ grows, where for large $N \gg 1$, $P_{\rm corr}$ saturates to 1.

In the intermediate monitoring strength regime $\gamma\in(0,1)$ where analytical control over the measurement distribution is limited, we numerically evaluate the likelihood of correctly inferring the state, defined as
\begin{equation}
P_{\rm{corr}}(\gamma) = \sum_{m, \psi\in\{\Psi_0, \Phi_0\}} p(\{m\}|\psi)p(\psi)p(\psi|\{m\}).
\label{eq:p_corr_n65}
\end{equation}
Substituting Eq.~\eqref{eq:pcorr_N}, we obtain the expression $P_{\rm{corr}}$ in terms of the set size $N$.
In Fig.~3(a) of the main text, we plot the steady-state likelihood $P_{\rm{corr}}$ for a fixed set size $N = 64$, as a function of measurement strength $\gamma$. This likelihood quantifies the ability to distinguish between two initial states $|\Psi_0\rangle$ and $|\Phi_0\rangle$, given a measurement trajectory ${m}$. We find that $P_{\rm{corr}}$ reaches unity at the critical point $\gamma_c = 0.39$, for system size $L =12$.

\section{Data acquisition}\label{sec:C}
In this appendix, we describe how we simulate the monitored circuits and obtain data for the two tasks discussed in the main text, namely state distinguishing and phase recognition. For both tasks, we consider circuits with a similar brickwork structure under weak measurements. We first present this general setup in SM section~\ref{sm:general_circuit_simulation}. Then, we discuss details specific to each task in SM sections~\ref{sec:state_distinguishability_circuit_simulation} and \ref{sec:phase_recognition_circuit_simulation} for state distinguishing and phase recognition, respectively. 
We then discuss how the baseline critical measurement strength $\gamma_c$ is estimated in SM section~\ref{sm:baseline_for_phase_recognition}. At the end of this section, we show insights into some statistical artifacts in the data (see SM Section~\ref{sec:C3}). 

\subsection{General Setup for Monitored Circuit Dynamics}\label{sm:general_circuit_simulation}
To investigate unitary-monitored dynamics, we consider a (1+1)d circuit generating hybrid dynamics acting on some initial state $|\psi\rangle$. 

The initial state $|\psi\rangle$ is an entangled and scrambled quantum state generated via applying unitary circuits to a chosen product state. The unitary circuits have a structure of interlacing layers of local two-site unitary gates applied alternately on even and odd bonds at each time step, commonly referred to as the brickwork structure. Defining one time step as one even layer followed by one odd layer, we evolve the product state for $t_s=2L$, where $L$ is the system size.  
To completely specify the initial state $|\psi\rangle$ used in each task, we specify the choices of the product state and the two-qubit gates used in the initial state preparation in the task-specific subsections below. 

The (1+1)d hybrid circuit dynamics consist of unitary evolution and mid-circuit measurements. For the unitary evolution, we again consider unitary circuits with a brickwork structure, where the evolution time is $T=2L$. 
These unitary circuits serve as minimal models for studying chaotic dynamics in closed quantum systems. 
From an information perspective, they tend to scramble local information into non-local degrees of freedom, effectively protecting the encoded state against local errors. 
The other component of hybrid dynamics is the mid-circuit measurements. To have control over the strength of monitoring, we consider a single-qubit weak measurement protocol that couples one system qubit to an ancilla initialized in state $|0\rangle$ by a varying degree $\gamma$ via a two-qubit unitary $U_c(\gamma)$, and then projectively measures the ancilla qubit. We illustrate the circuits underlying the weak measurement protocol in SFig.~\ref{fig:hybridsetup}. All physical qubits are measured through this single-qubit weak measurement protocol between consecutive unitary evolution layers, resulting in a measurement trajectory at a given time step. 
These measurements serve as local interactions that extract information from the environment, leading to non-unitary evolution. 

The resulting brickwork circuit structure with mid-circuit weak measurements is shown in SFig.~\ref{fig:hybridsetup}. The measurement trajectories are binary bitstrings with shape $L\times T=L\times 2L$. To completely specify the monitored circuits generating the hybrid dynamics used for each task, we specify the entangling gate $U_c(\gamma)$ used in weak measurement and the two-qubit gates used in the brickwork unitary circuit in the following task-specific subsections. 

For both the state distinguishing and phase recognition, we simulate the initial state $|\psi\rangle$ evolved by the hybrid circuit dynamics. The specific initial state(s) $|\psi\rangle$ and the choice of two-qubit gates in the brickwork structure differ for the two tasks; we provide details below. 

To generate the training data, we classically simulate our circuit setup to sample measurement trajectories ($\vec{m}$) across different circuit runs. These trajectories are then post-processed using QuAN to perform the phase classification task. 
For clarity, we highlight the most experiment-friendly setup that is accessible with current quantum hardware, using only translationally invariant gates during scrambling and monitored dynamics. To simulate the circuit in realistic near-term quantum hardwares, we use a depolarizing noise model with single-qubit and two-qubit gate error rates being $p_{1\rm{q}} = 4 \times 10^{-5}$ and $p_{1\rm{q}} = 2 \times 10^{-3}$, respectively.

\subsection{State Distinguishability Setup}\label{sec:state_distinguishability_circuit_simulation}

\begin{figure}[ht]
    \centering
    \includegraphics[page=2, width=0.95\linewidth]{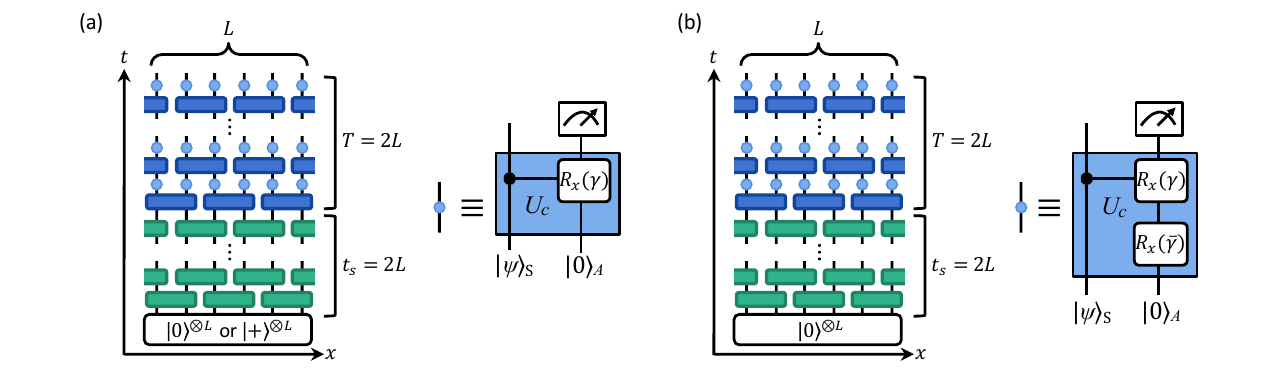}
    \caption{\textbf{Circuits for data generation:}  
    (a) Left: A sketch of the circuit setup for the state distinguishibility task. We randomly prepare one of the two initial states by evolving product states $|0\rangle^{\otimes L}$ and $|+\rangle^{\otimes L}$ (where $|+\rangle=\frac{1}{\sqrt{2}}(|0\rangle+|1\rangle)$) with a brickwork Harr random unitary scrambling circuit $U_s$ (green boxes). We then evolve the initial state with monitored circuit dynamics (blue) composed of unitary gate evolution $U_m$ (blue box) and weak measurements (blue dots). Right: A zoomed-in view of the weak measurement protocol used for state distinguishing. One ancilla qubit ($|0\rangle_A$) is entangled with a system qubit $|\psi\rangle_S$ through the entangling operation $U_c$ (blue box). The ancilla qubit is then projectively measured. 
    (b) Left: A sketch of the circuit setup for the phase recognition task. Right: A zoomed-in view of the weak measurement protocol used in phase recognition. Note the differences in the choice of product state in initial state preparation and the details inside the weak measurement protocol. 
    }
    \label{fig:hybridsetup}
\end{figure} 

In this subsection, we provide details about parameters and setups specific to the state-distinguishing task (see SFig.~\ref{fig:hybridsetup}(a)). 

First, we specify parameters used for initial state preparation. For state distinguishability, we consider two initial states $|\psi\rangle \in \{|\Psi_0\rangle, |\Phi_0\rangle\}$ with system size $L$, which are reached via applying the same brickwork unitary circuit onto product states $|0\rangle^{\otimes L}$ and $|+\rangle^{\otimes L}$ ($|+\rangle=\frac{1}{\sqrt{2}}\left(|0\rangle+|1\rangle\right)$).
The two-qubit gates in the brickwork unitary are the same Haar unitary, whose exact form is as follows:
\begin{equation} \label{eq:state_distinguishing_us}
 U_{s} = 
\begin{bmatrix}
0.3644 + 0.3086i & 0.2537 + 0.0937i & 0.5589 + 0.0768i & 0.5589 - 0.2612i \\
0.3857 - 0.5273i & -0.1871 + 0.1649i & 0.3448 + 0.5512i & -0.2860 + 0.0803i \\
-0.0213 + 0.4688i & 0.0841 - 0.4706i & 0.4436 + 0.0357i & -0.5604 + 0.1975i \\
0.1572 + 0.3166i & 0.0522 + 0.7958i & 0.0076 - 0.2467i & -0.4211 - 0.0276i
\end{bmatrix}.
\end{equation}
We randomly select a state from $\{|\Psi_0\rangle, |\Phi_0\rangle\}$ with equal probability and evolve it with the hybrid circuit dynamics.

Second, we specify the parameters used in the hybrid dynamics circuit. 
The two-qubit gates that drive unitary evolution in the hybrid circuit dynamics are the same as the ones used in the initial state preparation, $U_m=U_s$, where $U_s$ is defined in Eq.~\eqref{eq:state_distinguishing_us}. 
For the weak measurement protocol, $U_c(\gamma)$ is given by a controlled rotation gate $\rm{CR}_{x}(\gamma)$ where the system qubit controls the operation on the target ancilla (see SFig.~\ref{fig:hybridsetup}(a)). The exact form of the $U_c(\gamma)$ gate is given by
\begin{equation}
    U_c(\gamma) = \rm{CR}_{x}(\gamma) = |0\rangle\langle0| \otimes \rm{I} +  |1\rangle\langle 1| \otimes \exp\left(\frac{i\pi\gamma\sigma_{x}}{2}\right).
    \label{eq:state_distuinguishing_gate}
\end{equation}
The rotation angle $\gamma$ controls the weak measurement strength. Finally, the ancilla qubit is measured in its computational basis to obtain a binary measurement outcome: $0$ or $1$. 
In practice, this single-qubit weak measurement protocol only uses one ancilla qubit so that the same ancilla qubit can be reused to measure all the physical qubits one at a time. 

Here, we further discuss some intuitive interpretation of the measurement strength $\gamma$ with the given form of $U_c$ in Eq.~\eqref{eq:state_distuinguishing_gate}, which varies between $0$ and $1$. 
At the lower limit ($\gamma = 0$), the ancilla is left decoupled from the system. Therefore, every measurement deterministically yields the same outcome $0$ at all sites for both initial states. 
Meanwhile, in the other limit of $\gamma = 1$, the measurements of system qubits occur projectively in the $\sigma_{z}$ basis. This, in turn, leads to most information about the state being contained only at early times in the ancilla measurement outcomes. For a $\gamma$ between $[0,1]$, the measurement outcomes recover only partial information about the system qubit. We input these measurement trajectories $\vec{m}$, along with its correct state label ($|\Psi_{0}\rangle$ or $|\Phi_{0}\rangle$), to achieve the supervised training of QuAN.

\subsection{Phase Recognition Setup}\label{sec:phase_recognition_circuit_simulation}

The phase-recognition setup hinges on learning the difference in crucial characteristics between the mixed and pure phases of MIPT. 
We made adequate changes in our protocol, which takes into account the distinct nature of the classification task compared to the state distinguishability setup (see SFig.~\ref{fig:hybridsetup}(b)). 

For phase recognition, we only consider one initial state. The initial state $|\Psi_0\rangle$ is generated by evolving the product state $|0\rangle^{\otimes L}$ using the brickwork circuit composed of the scrambling Haar unitary $U_s$ (Eq.~\eqref{eq:state_distinguishing_us}). The scrambling circuit used to generate the initial state is the same as the one used in state distinguishability. 

We then evolve the scrambled initial state with monitored circuit dynamics.  
The two-qubit gates in the brickwork unitary evolution part of the monitored dynamics are given by
\begin{equation}
 U_{m} = 
\begin{bmatrix}
0.9167 - 0.1057i & 0.3727 + 0.0430i & -0.0300 - 0.0692i & -0.0181 + 0.0419i \\
0.0188 - 0.0022i & 0.1810 + 0.0209i & 0.3601 + 0.8311i & 0.1519 - 0.3507i \\
-0.0438 - 0.3797i & -0.1037 + 0.8998i & 0.1672 - 0.0724i & 0.0174 + 0.0075i \\
0.0052 + 0.0454i & 0.0086 - 0.0750i & 0.3443 - 0.1491i & 0.8467 + 0.3668i
\end{bmatrix},
\label{eq:monitored_gate}
\end{equation}
where the choice of unitary $U_{m}$ is made to strengthen the measurement dynamics over the scrambling dynamics in the monitored layers. 
The unitary coupling gate $U_{c}(\gamma)$ used in the weak measurement protocol is given by 
\begin{equation}
    U_{c}(\gamma) = \rm{CR}_{x}(\gamma) \cdot (\rm{I}\otimes R_{x}(\bar{\gamma}))  =|0\rangle\langle0| \otimes \exp\left(\frac{i\pi(1-\gamma)\sigma_{x}}{4}\right) +  |1\rangle\langle 1| \otimes \exp\left(\frac{i\pi(1+\gamma)\sigma_{x}}{4}\right),
    \label{eq:coupling_gate}
\end{equation}
where $R_{x}(\bar{\gamma})$ is a single qubit rotation applied first to the ancilla qubit, with rotation angle $\bar{\gamma} = \frac{\pi(1-\gamma)}{2}$. 
We note that the controlled rotation gate $\rm{CR}x(\gamma)$ is the same as the one used for the state distinguishability setup. The difference here is that the additional rotation gate $R_{x}(\bar{\gamma})$ applied to the ancilla qubit prevents any apparent statistical signal in the measurement outcomes, such as the density of having outcome $0$, while still ensuring that the two extreme $\gamma$ limits, $\gamma=0$ and $\gamma=1$, correspond to no measurement and measuring the system qubit projectively at all times, respectively. $U_c(\gamma=0)$ for the phase recognition setup corresponds to having the ancilla initialized in the $\sigma_y=1$ eigenstate, then entangle with the system qubit. For $\gamma=1$, Eq.~\eqref{eq:coupling_gate} restore back to Eq.~\eqref{eq:state_distuinguishing_gate} in state distinguishing. 

For the phase recognition task, we input measurement trajectories $\vec{m}$, along with assigned labels depending on the measurement strength $\gamma$ used to generate  $\vec{m}$, to achieve the supervised training of QuAN.

\begin{figure}[ht]
    \centering
    \includegraphics[page=3, width=0.95\linewidth]{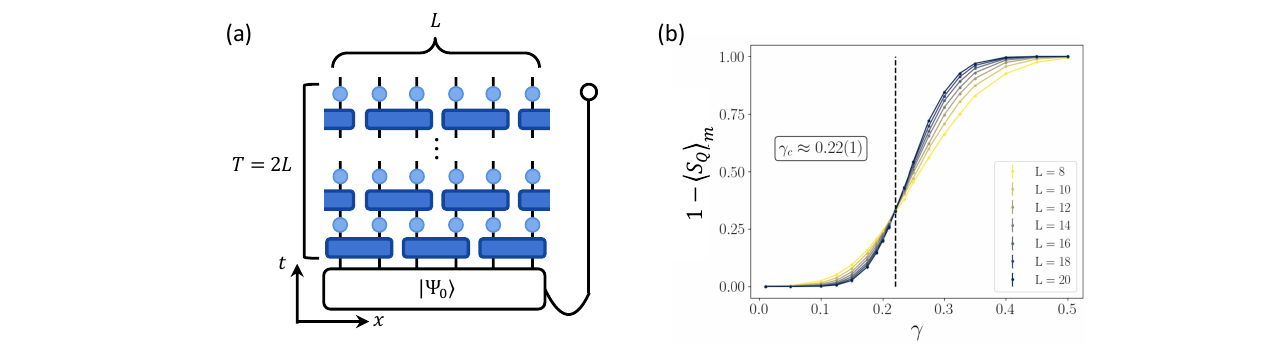}
    \caption{\textbf{Benchmarking phase recognition:}  (a) 
    A sketch of the circuit used to identify the baseline critical measurement strength. Similar to the setup shown in SFig.~\ref{fig:hybridsetup}(b) except an extra reference qubit $(Q)$ (black circle) that is coupled to one of the system qubits by forming a maximally entangled Bell state at $t = 0$.
    (b) We plot $ 1 - \langle S_{Q} \rangle_m$ against the measurement strength $\gamma$, where unitary dynamics keep the reference qubit coupled with all the system qubits, giving $S_{Q} = 1$ at $\gamma = 0$. In contrast, for $\gamma =1$, the projective measurements completely decouple the reference qubit $(Q)$ from the rest of the system and lead to a product state $S_{Q} = 0$. Each data point is averaged over $M=25,000$ trajectories, and a finite-size scaling analysis of $\langle{S}_{Q}\rangle_m$ shows a clear crossing, marking the transition point $\gamma_{c} = 0.22(1)$.
    }
    \label{fig:ancilla_op}
\end{figure} 

\subsection{Baseline for phase recognition}
\label{sm:baseline_for_phase_recognition}
To benchmark QuAN's prediction of the transition point, we estimate the critical measurement strength through a different protocol. 
We consider a slightly different circuit with an extra reference qubit (Q) maximally entangled to one of the system qubits in the product state used in initial state preparation, as shown in SFig.~\ref{fig:ancilla_op}~\cite{gullans2020Phys.Rev.Lett.}. All other setups for the initial state preparation and hybrid circuit dynamics are the same as the ones used in phase recognition. 
The critical measurement strength can be identified through a finite-size scaling analysis of Von Neumann entanglement entropy of a reference qubit (Q), defined as 
\begin{equation}
    \langle S_Q\rangle _m = -\frac{1}{M}\sum_i^M \text{Tr} (\rho_{Q,i} \log_{2} \rho_{Q,i})
\end{equation}
where $\rho_{Q,i}$ is the reduced density matrix of the reference qubit $Q$ corresponding to the $i$-th trajectory out of $M$ total trajectories.
As showing in SFig.~\ref{fig:ancilla_op}(b)), the critical measurement strength is at  $\gamma_{c} = 0.22(1)$~\cite{szyniszewski2019entanglement,bao2020Phys.Rev.B,agrawal2024Phys.Rev.X,aziz2024critical}. 

\subsection{Spatiotemporal correlations in the measurement record as finite-size artifacts}\label{sec:C3}
\begin{figure}[ht]
    \centering
    \includegraphics[page=4, width=0.9\linewidth]{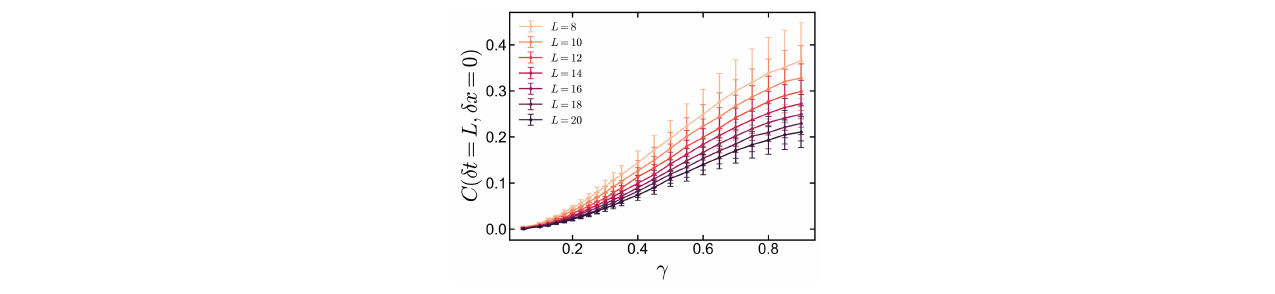}
    \caption{Spatio-temporal correlation $C(\delta t=L, \delta x=0)$ as a function of measurement strength $\gamma$ for varying system size $L$. The error bar represents the standard error from averaging over different spacetime points $t_0, x_0$. We use $M=10,000$ measurement outcomes to compute the correlation function. }
    \label{fig:tcorr}
\end{figure}
In this section, we examine the spatiotemporal correlation in the measurement outcome to better understand the spurious signatures within the measurement record. 
These correlations are finite-size artifacts resulting from the periodic boundary condition and the brick-wall circuit dynamics.
The average correlation function between two measurement outcomes separated in time $\delta t$ at equal spatial position can be defined as 
\begin{align}
C(\delta t, \delta x=0) =& \frac{1}{L}\frac{1}{T-\delta t}\sum_{x_0=1}^L\sum_{t_0=1}^{T-\delta t} C(t_0,t_0+\delta t,x_0,x_0) \\
=& \frac{1}{L}\frac{1}{T-\delta t}\sum_{x_0=1}^L\sum_{t_0=1}^{T-\delta t} \left\{ \frac{1}{M}\sum_{i=1}^M m_i(t_0,x_0)m_i(t_0+\delta t,x_0)\right\}, \label{eq:C1}
\end{align}
where $m_i(t,x)$ denotes the binary measurement outcome at spacetime point $(t,x)$ in trajectory $i$, and $M$ is the total number of trajectories. 
When time-separation hits the system size, i.e., $\delta t=L$, the signal starting at $(x_0, t_0)$ can wrap around the system and reach back to the same site at $(x_0, t_0 + L)$ in a system of size $L$, causing a temporal correlation between $m_i(t_0,x_0)$ and $m_i(t_0+L,x_0)$. 
This finite-size effect will become more prominent with increasing monitoring strength $\gamma$,
where measurement outcomes more faithfully reflect the system’s internal state. 
As shown in SFig.~\ref{fig:tcorr}, the correlation $C(\delta t = L, \delta x = 0)$ increases gradually with $\gamma$. The gradual nature of this increase contrasts with the sharp onset observed in phase recognition, implying that QuAN's learning of phase recognition is not driven by learning this spurious correlation.

\section{Estimation of \texorpdfstring{$\Prob(p)$}{Prob(p)} from the measurement record}\label{sec:D}

In this section, we describe the procedure for estimating the distribution $\Prob(p)$ of Born probabilities from the measurement record, under the assumption that the underlying true Born probabilities are unknown and must be inferred from data.

We begin by approximating the Born probability of a specific measurement outcome $m(t)$ at a fixed time using its empirical frequency across $M$ trajectories,
\begin{align}
    p(m)_t = \frac{k_{m(t)}}{M}
\end{align}
where $k(m(t))$ is the number of trajectories that produced the specific outcome string $m(t)$ at time $t$.
To account for the periodic boundary condition in the spatial dimension, where trajectories are defined on a cylinder, we treat measurement outcome strings related by spatial translation as physically equivalent. For a system of size $L$, each trajectory is translated by $L$ shifts, effectively generating $L$ spatially shifted versions of each measurement string. These translations preserve the physical content due to periodicity and enhance the statistical resolution of the probability estimate.

Using the estimated Born probabilities, we construct a histogram of those discrete probabilities $p(m)_t$ as an estimate of the distribution of Born probability $\Prob(p)$,
\begin{align}
    \Prob(p) = \frac{N_p}{D}
\end{align}
where $D=2^L$ is the Hilbert space size for $L$ system qubits, and $N_p$ represents the number of distinct $m(t)$ that has same Born probability value $p(m)_t$. This provides an estimate of the distribution $\Prob(p)$ that captures the frequency distribution of Born probabilities.

\subsection{Time-independence of Born probabilitiy \texorpdfstring{$p(\vec{m})_t$}{p(m)t} in weak monitoring phase}\label{sec:D1}

\begin{figure}[ht]
    \centering
    \includegraphics[page=15, width=0.9\linewidth]{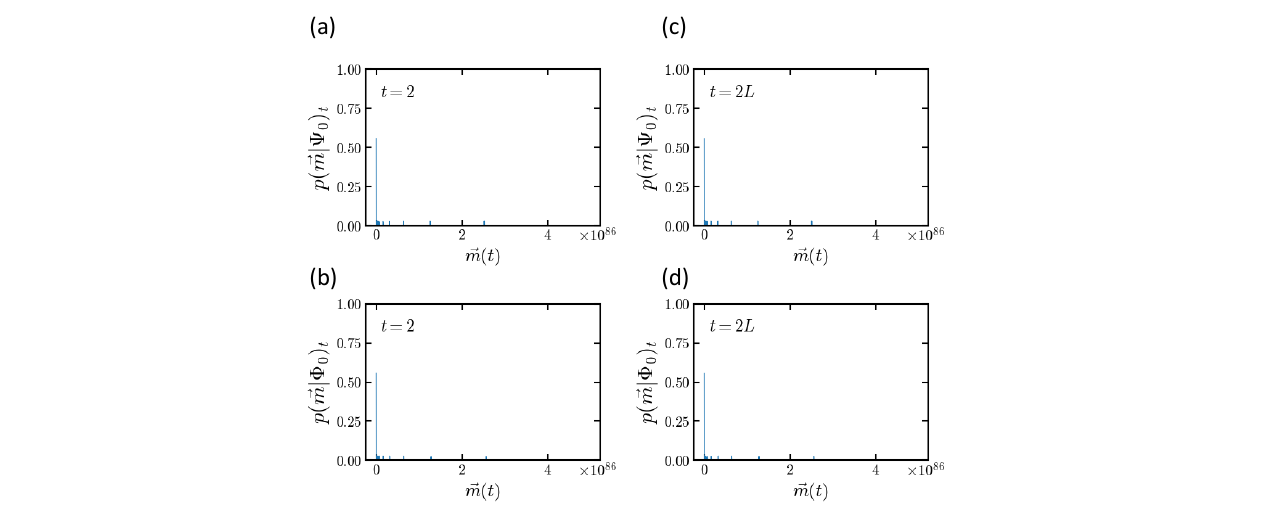}
    \caption{Estimated Born probabilities of measurement outcome at fixed time $t$ for small $\gamma=0.2$. (a,b) Estimated Born probabilities at early time $t=2$ from initial states (a) $\ket{\Psi_0}$ and (b) $\ket{\Phi_0}$. (c,d) Estimated Born probabilities at late time $t=2L$ for initial state (c) $\ket{\Psi_0}$ and (d) $\ket{\Phi_0}$.}
    \label{fig:3}
\end{figure}

This subsection complements the claim made in the main text: under strong monitoring, the initial states can be distinguished at early times, whereas under weak monitoring, the measurement outcomes fail to reflect the system’s initial state at any time. While strong monitoring enables early-time measurements to capture distinct signatures of different initial states, we demonstrate here that this distinction is absent in the weak monitoring regime. Specifically, we examine the Born probability $p(\vec{m})_t$ at fixed times for weak measurement strength $\gamma = 0.2$ and demonstrate that the probabilities are nearly identical for different initial states and remain time independent.

To this end, we use $M = 6,000,000$ measurement samples to estimate $p(\vec{m})_t$ for both initial states $\ket{\Psi_0}$ and $\ket{\Phi_0}$ at weak monitoring strength $\gamma = 0.2$, as shown in SFig.~\ref{fig:3}.
SFig.~\ref{fig:3}(a,b) shows the Born probability at early time $t = 2$ and (c,d) at late time $t = 2L$. Across all time slices, the probabilities corresponding to the two initial states are nearly indistinguishable. This confirms that, in the weak monitoring regime, the ancilla measurements are insensitive to the system’s initial state, and the distribution $p(\vec{m}|\psi)_t$ remains effectively time-independent.
As discussed in the main text, this time-independence at small $\gamma$ highlights that the system retains no usable information about its initial state within the measurement record at any fixed time, failing to distinguish between two initial states.

\section{Model Architecture and Training scheme}\label{sec:E}
\subsection{QuAN architecture}\label{sec:E1}
We provide a detailed description of our architecture, which builds upon the QuAN's design~\cite{kim2024}. A schematic architecture is shown in Fig.~2(d) of the main text. 
Our discussion focuses on the newly added modules compared to the original QuAN framework.
First, we discuss the input data structure and embedding layer, which are designed for brickwork structured data. Next, we discuss an encoder consisting of a temporal attention and inter-trajectory attention, mainly focusing on the usage of temporal attention to inspect the temporal structure of the data. Lastly, we discuss the pooling attention block (PAB) decoder.

\subsubsection{Embedding}

The architecture processes a set of binary-valued measurement outcomes $\mathbb{m}=\{m^\alpha_{tx}\}_{\alpha=1}^{N}$, where $m^\alpha_{tx}$ denotes a trajectory of (1+1)-dimensional binary array (0, 1). Each trajectory contains $N_q=2L^2$ binary values. 
The superscript index $\alpha$ runs from $1$ to $N$, labeling individual trajectories within the set. The subscript indices $t$ and $x$ correspond to time and space, with $t=1,\cdots,2L$ and $x=1,\cdots,L$. For simplicity, we refer to measurement outcomes at the same time step as a `single time slice'. (e.g. in $m^\alpha_{tx}$, there are $2L$ time slices, and each time slices consist of $L$ binary values.)

A set of trajectories $\mathbb{m}$ goes through the embedding layer $\texttt{emb}$. (See Fig.~2(c) in the main text.) Embedding layer transforms binary values into real values of continuous entries to ensure better algebraic properties, while capturing temporal features specific to brickwork structured data through the following steps: 
First, neighboring measurement outcomes are grouped into clusters based on the underlying brickwork structure of the monitoring circuit (see main text Fig.~2(c)). Each cluster consists of two binary outcomes (four different kinds of clusters exist - $00$, $01$, $10$, and $11$), resulting in $L/2$ clusters within a single time slice.
Next, these clustered binary values are mapped to a real-valued embedded vector ($\vec{v}_{00}$, $\vec{v}_{01}$, $\vec{v}_{10}$, and $\vec{v}_{11}$). This embedding operation can be expressed as $\mathbb{Z}_2^2\to\mathbb{R}^{n_e}$, where the length of the embedded vector $n_e$ is a model hyperparameter. 
Finally, two consecutive time slices (odd and even) are combined into a single new time slice, indexed by $\tau$. The embedded vectors within each new time slice are concatenated, forming a flattened single vector representation. As a result, the output of the embedding layer,  $\mathbb{x}\equiv\{\mathbb{x}^\alpha_{\tau\mu}\}_{\alpha=1}^N=\{\texttt{emb}(m^\alpha_{tx})\}_{\alpha=1}^N$ is a set of (1+1)-dimensional arrays with new time index $\tau=1,\cdots,L$ and new space index $\mu=1,\cdots,n_eL$.

\subsubsection{Encoder}
The original purpose of using QuAN's encoder with a self-attention block was to inspect high-order moments of the bitstring distribution and effectively learn the statistics of the distribution~\cite{kim2024}. 
However, unlike static quantum measurement data, which all originate from the final state measurement, we work with dynamic quantum data, where the temporal dimension plays a significant role. 
To account for this, we introduce a new temporal attention module designed to capture the temporal structure of the measurement outcomes. 

\paragraph{Temporal attention.}
The input to the encoder is denoted as $\mathbb{x}^\alpha_{\tau\mu}$, where $\alpha$ indexes the set element and $\tau$ and $\mu$ index the time and space dimension of the feature space after the embedding layer. Here, we use Greek letters $\tau, \sigma$ to represent the time index and $\mu, \nu, \rho, \kappa, \lambda$ for the space index. First, the temporal attention transforms input time slices to a hidden-state-vector time slice:
\begin{equation}\label{eq:temporalattn}
    \mathbb{h}^\alpha_{\tau\mu} = \sum_{\nu}^{n_eL}Q^{(t)}_{\mu\nu}\mathbb{x}^\alpha_{\tau\nu} + \sum_{\sigma=1}^{L}\texttt{Softmax}\left[\sum_\rho^{d_h}\sum_{\kappa\lambda}^{ne_L}\frac{1}{\sqrt{d_h}}Q^{(t)}_{\rho\kappa}\mathbb{x}^\alpha_{\tau\kappa}K^{(t)}_{\rho\lambda}\mathbb{x}^\alpha_{\sigma\lambda}\right]\sum_\nu^{d_x}V^{(t)}_{\mu\nu}\mathbb{x}^\alpha_{\sigma\nu},
\end{equation}
with the hidden dimension size $d_h$. $Q^{(t)}, K^{(t)}, V^{(t)}$ are query, key, value matrices of dimension $(d_h, d_x)$. The temporal attention score of size $L\times L$ is given by 
\begin{align}\label{eq:temporalattentionscore}
     \texttt{Softmax}[A^{(t)}(\tau, \sigma)] = \texttt{Softmax}\left[\sum_\rho^{d_h}\sum_{\kappa\lambda}^{ne_L}\frac{1}{\sqrt{d_h}}Q^{(t)}_{\rho\kappa}\mathbb{x}^\alpha_{\tau\kappa}K^{(t)}_{\rho\lambda}\mathbb{x}^\alpha_{\sigma\lambda}\right],
\end{align}
with normalization $\sum_\sigma A_{\tau\sigma}=1$.
Next, layer normalization on the space dimension and a linear layer follow:
\begin{align}
    \mathbb{h}'^\alpha_{\tau\mu} &= \texttt{LayerNorm}(\mathbb{h}^\alpha_{\tau\mu}) \\
    \texttt{SAB}^{(t)}(\mathbb{x}^\alpha_{\tau\mu}) &= \texttt{LayerNorm}(\mathbb{h}'^\alpha_{\tau\mu}+\texttt{FF}_{\mu\nu}(\mathbb{h}'^\alpha_{\tau\nu})),
\end{align}
where $\texttt{FF}(\mathbb{x}) = \texttt{ReLU}(O^{(t)}\mathbb{x})$ is a feed-forward function for residual connection that acts on each time slice equally by multiplying the matrix $O^{(t)}$ of dimension $(d_h, d_h)$ followed by the activation function $\texttt{ReLU}$.
One important thing to notice is that temporal attention $\texttt{SAB}^{(t)}$ only inspects intra-set correlation, which is the temporal correlation within a single trajectory. 
We use two layers of $\texttt{SAB}^{(t)}$ to inspect fourth-order temporal correlation. 
\begin{align}
    \mathbb{x}'^\alpha_{\tau\mu} &= \texttt{Sigmoid}\left(\texttt{SAB}^{(t)} \circ \texttt{SAB}^{(t)}(\mathbb{x}^\alpha_{\tau\mu})\right) \\
    \mathbb{y}^\alpha_\tau &\equiv \texttt{T-Attn}(\mathbb{x}^\alpha_{\tau\mu})= \sum_\mu W^{(t)}_\mu \texttt{LayerNorm}(\texttt{FF}_{\mu\nu}(\mathbb{x}'^\alpha_{\tau\nu})) + b^{(t)}.
\end{align} 
The linear layer $W^{(t)}$ is a matrix of dimension $(1, d_h)$ that converts each time slice $\mathbb{x}'^\alpha_{\tau\mu}$ into scalar value $\mathbb{y}^\alpha_\tau$, outputting single vector of length $L$ for each trajectories. 
Therefore, the output of t-Transf. $\mathbb{y}^\alpha_\tau\equiv \texttt{T-Attn}(\mathbb{x}^\alpha_{\tau\mu})$ is a matrix of dimension $(N, L)$.

\paragraph{Inter-trajectory attention.}
Now we move on to the inter-trajectory attention that utilizes a self-attention block (SAB). We use the same structure as introduced in Ref.~\cite{kim2024}. SAB transforms the output set of temporal attention, $\mathbb{y}^\alpha_\tau$, to a set of hidden state vectors:
\begin{align}
    \mathbb{g}^\alpha_\tau &= \sum_\sigma^L Q^{(s)}_{\tau\sigma} \mathbb{y}^\alpha_\sigma + \sum_{\beta=1}^N \texttt{Softmax}\left[\sum_\eta^L \sum_{\zeta\xi}^L\frac{1}{\sqrt{L}} Q^{(s)}_{\eta\zeta} \mathbb{y}^\alpha_\zeta K^{(s)}_{\eta\xi} \mathbb{y}^\beta_\xi \right] \sum_\sigma^L V^{(s)}_{\tau\sigma} \mathbb{y}^\beta_\sigma \\
    \mathbb{g}'^\alpha_\tau &= \texttt{LayerNorm}(\mathbb{g}^\alpha_\tau) \\
    \texttt{SAB}(\mathbb{y}^\alpha_\tau) &= \texttt{LayerNorm}(\mathbb{g}'^\alpha_\tau + \texttt{FF}_{\tau\sigma}(\mathbb{g}'^\alpha_\sigma)),
\end{align}
where $Q^{(s)}, K^{(s)}, V^{(s)}$ are query, key, value matrices of the self-attention block. 
The inter-trajectory attention score of size $N\times N$ is given by 
\begin{align}\label{eq:intertrajattentionscore}
    \texttt{Softmax}[A(m_\alpha,m_\beta)] = \texttt{Softmax}\left[\sum_\eta^L \sum_{\zeta\xi}^L\frac{1}{\sqrt{L}} Q^{(s)}_{\eta\zeta} \mathbb{y}^\alpha_\zeta K^{(s)}_{\eta\xi} \mathbb{y}^\beta_\xi \right].
\end{align}
Each element in the inter-trajectory attention score involves two set elements ($\mathbb{y}^\alpha$ and $\mathbb{y}^\beta$), which can capture all-to-all second-order moments in $\mathbb{y}$. We use two layers of $\texttt{SAB}$ to inspect 4th order moments of the trajectory distribution. The output of these SAB layers is given by
\begin{equation}
    \mathbb{z}^\alpha_\tau \equiv \texttt{InterTraj-Attn}(\mathbb{y}^\alpha_{\tau}) = \texttt{Sigmoid}(\texttt{SAB}\circ\texttt{SAB}(\mathbb{y}^\alpha_{\tau})),
\end{equation}
where $\mathbb{z}^\alpha_\tau\equiv \texttt{InterTraj-Attn}(\mathbb{y}^\alpha_{\tau}) $ is a matrix of dimension $(N, L)$.

\subsubsection{Pooling Attention Block (PAB) Decoder}
We use the same decoder structure as Ref.~\cite{kim2024} to respect the permutation invariance of the trajectories, while enabling importance sampling by assigning weights through pooling attention scores. 
\begin{align}
    \mathbb{p}_\tau &= S^{(p)} + \sum_{\alpha=1}^N\texttt{Softmax}\left[\sum_{\eta\zeta}^L \frac{1}{\sqrt{L}}S^{(p)}_\eta K^{(p)}_{\eta\zeta}\mathbb{z}^\alpha\zeta\right]\sum_\sigma^L V^{(P)}_{\tau\sigma}\mathbb{z}^\alpha\sigma \\
    \mathbb{p}'_\tau &= \texttt{LayerNorm}(\mathbb{p}_\tau) \\
    y(\mathbb{m}) &= \texttt{Sigmoid}\left(\sum_\tau W^{(p)}_\tau \texttt{LayerNorm}(\mathbb{p}'_\tau + \texttt{FF}_{\tau\sigma}(\mathbb{p}'_\sigma))+b^{(p)}\right),
\end{align}
where $S^{(p)}$ is a seed vector which is used as a query vector of size $(1, L)$ for a weighted average $\mathbb{z}^\beta$ over the set dimension. $K^{(p)}, V^{(p)}$ are key, value matrices of self-attention block. $W^{(p)}$ is a matrix of dimension $(1, L)$ that converts the vector $\mathbb{p}'_\tau$ into a single scalar output $y$, which is the confidence of predicting given set $\mathbb{m}$ into strong monitoring phase.

\subsection{Data preprocessing, training, and testing scheme}\label{sec:E2}

\begin{table}[ht] \setlength{\tabcolsep}{8pt}
\begin{tabular}{|l|c|}
\hline
\multicolumn{2}{|l|}{Model hyperparameter} \\ \hline
Embedding layer hyperparameter ($n_e$)  & 4      \\
Set size ($N$)                          & $64$      \\
Number of tSAB, SAB layers              & 2      \\
tSAB: Hidden spatial dimension ($d_h$)  & 16     \\
tSAB: Attention score drop rate ($p_d$)   & $0.1$     \\
tSAB, SAB: Number of heads              & 1      \\\hline\hline
\multicolumn{2}{|l|}{Training hyperparameters for state distinguishing task}                       \\\hline
Loss function           & Binary cross entropy loss (\verb|BCELoss|) \\
Optimizer               & \verb|Adam|$(\beta_1 = 0.9, \;\beta_2 = 0.999, \;\epsilon = 1\times 10^{-8})$ \\
L2 coefficient          & $5\times10^{-5}$      \\
Learning rate           & $1\times10^{-4}$      \\
Epoch                   & $600,000,000/M$           \\
Dataset shuffling period& 10                    \\
Batchsize               & $2048/N$             \\
Initialization          & Default               \\ 
GPU                     & A100 (80GB)           \\\hline\multicolumn{2}{|l|}{Training hyperparameters for phase recognition task}                       \\\hline
Loss function           & Binary cross entropy loss (\verb|BCELoss|) \\
Optimizer               & \verb|Adam|$(\beta_1 = 0.9, \;\beta_2 = 0.999, \;\epsilon = 1\times 10^{-8})$ \\
L2 coefficient          & $5\times10^{-5}$      \\
Learning rate           & $1\times10^{-4}$      \\
Epoch                   & $8000000/M$           \\
Dataset shuffling period& 10                    \\
Batchsize               & $32768/N$             \\
Initialization          & Default               \\ 
GPU                     & A100 (80GB)           \\\hline
\end{tabular}
\caption{Model setting and training hyperparameters.}
\label{tab:details}
\end{table}

We use PyTorch to train and test the model by minimizing the binary cross-entropy loss function $\mathcal{L}_{\rm BCE}$ of the true outputs and machine-predicted outputs $y(\mathbb{m})$, using the Adam optimization algorithm. 
To prevent the model from overfitting the data, we employ several methods. First, we shuffle the input sets every 10 epochs, ensuring the model explores various combinations of measurement outcomes~\cite{kim2024}. 
Second, we utilize DropAttention~\cite{zehui2019} in temporal attention, which enhances performance and reduces overfitting in Transformer architectures. Specifically, we randomly drop 10\% ($p=0.1$) of the attention score elements in the attention weight matrix during training. This dropout layer is discarded during testing. Other model/training hyperparameters are listed in Table~\ref{tab:details}. We train each set size independently while keeping other model parameters unchanged. 

For the state distinguishing task (see main text Fig.~3(a)), the training loss is given by 
\begin{align}\label{eq:bceloss1}
    \mathcal{L}_{\rm BCE} = -\frac{1}{\lfloor 2M/N \rfloor}\sum_i [Y_i \log y(\mathbb{m}_i) + (1-Y_i) \log (1-y(\mathbb{m}_i))],
\end{align}
where $Y_i=0$ for $\mathbb{m}_i$ from initial state $\Psi_0$ and $Y_i=1$ for $\Phi_0$. After training, we store the model with the lowest testing loss along the training epoch (denoted as `trained model' $y^*$), and use this trained model for evaluating the average likelihood of inferring the initial state $\overline{P_{\rm corr}|_{y^*}}$. The test loss is evaluated in the same way as the training loss, however, using the testing dataset.
We use $M=6,000,000$ trajectories for each initial state with system size of $L=12$ as a training dataset, and $M=1,000,000$ for a testing dataset. (The testing dataset size is fixed during the sample complexity study.) We construct a set $\mathbb{m}_i$ of set size $N$ by partitioning total trajectories into $\lfloor M/N \rfloor$ sets for each $\gamma$ and each initial state.

For the phase recognition task (see main text Fig.~4(a)), we employ the same model architecture and hyperparameters as in the state distinguishing task, using $M=832$ trajectories for a system size of $L=20$ to demonstrate performance. The training loss is given by
\begin{align}\label{eq:bceloss2}
    \mathcal{L}_{\rm BCE} = -\frac{1}{\lfloor 4M/N \rfloor}\sum_i [Y_i \log y(\mathbb{m}_{i,\gamma}) + (1-Y_i) \log (1-y(\mathbb{m}_{i,\gamma}))],
\end{align}
where $\mathbb{m}_{i,\gamma}$ denotes $i$-th set composed with trajectories from $\gamma$ and $Y_i=0$ for $\gamma\in\{0.05, 0.1\}$ and $Y_i=1$ for $\gamma\in\{0.85, 0.9\}$. Same as the state distinguishing task, we choose the model with the lowest testing loss along the training epoch, which is calculated over trajectories from 4 training $\gamma$ points ($\gamma=0.05, 0.1, 0.85, 0.9$, denoted as a star symbol in the main text Fig.~4(a)).
\begin{align}
    \mathcal{L}_{\rm BCE}|_{y^*}(M) = -\frac{1}{\lfloor 4M/N \rfloor}\sum_i [\log (1-y^*(\mathbb{m}_{i,\gamma=0.05})) + \log (1-y^*(\mathbb{m}_{i,\gamma=0.1})) + \log y^*(\mathbb{m}_{i,\gamma=0.85}) + \log y^*(\mathbb{m}_{i,\gamma=0.9})].
\end{align}
We use this trained model to evaluate the average strong monitoring phase prediction $\overline{\langle y^*\rangle}$.
To determine the appropriate number of training trajectories, we conduct a sample complexity study by scanning training sample size $M$ from $40,000$ to $256$ (see Appendix~\ref{sec:F3} for more details).  We use a fixed number of testing trajectories ($M=10,000$) throughout the sample complexity study to ensure that the model’s generalization ability is evaluated consistently, allowing us to identify the minimal training sample size required for successful generalization to large test sets.

\section{Extended machine learning results}\label{sec:F}

\subsection{Optimal hyperparameters study for state distinguishing task}\label{sec:F1}

\begin{figure}[ht]
    \centering
    \includegraphics[page=5, width=0.8\linewidth]{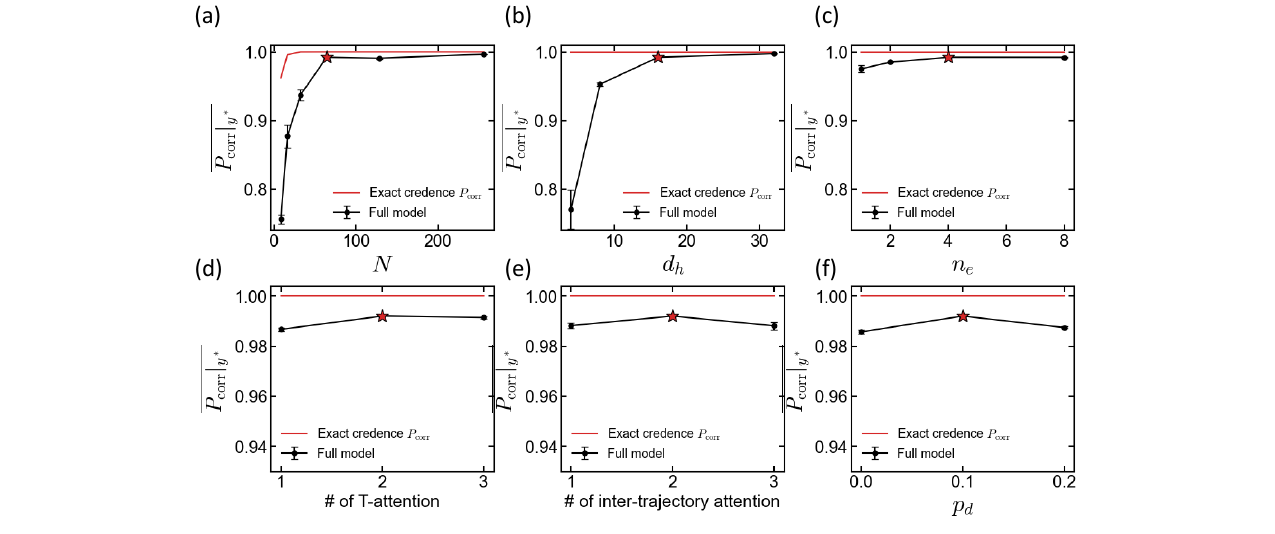}
    \caption{Optimal hyperparameter study for state distinguishing task using $L=12$ and $\gamma=0.9$. Average likelihood $\overline{P_{\rm corr}|_{y^*}}$ as a function of (a) set size $N$, (b) hidden dimension size $d_h$, (c) embedding vector size $n_e$, (d) number of temporal attention layer, (e) number of inter-trajectory attention layer, and (f) temporal attention score drop rate $p_d$. The red line represents the exact $P_{\rm corr}$, and the red star is the optimal likelihood we use in the main text. The error bar represents the standard error over 4 independently trained models.}
    \label{fig:6}
\end{figure}

In SFig.~\ref{fig:6}, we present the set size optimization study as well as the model hyperparameter tuning for the state distinguishing task.
First, we train QuAN on noiseless data from system size $L=20$ and $\gamma=0.9$, varying the set size $N = 8, 16, 32, 64, 128$, and $256$. We plot the sample mean of average likelihood $\overline{P_{\rm corr}|_{y^*}}$ over 4 independent trained models as a function of set size $N$, alongside the exact likelihood (see SFig.~\ref{fig:6}(a)).
Both the exact and QuAN-estimated $\overline{P_{\rm corr}|_{y^*}}$ increase with set size, but the exact likelihood saturates more quickly than QuAN’s estimate. Based on the set size at which QuAN’s likelihood saturates near 1, we conclude that the optimal set size is $N_{\mathrm{op}} = 64$.
In SFig.~\ref{fig:6}(b–f), we conduct a hyperparameter study using the same dataset and the optimal set size. The hyperparameters considered include the hidden dimension size $d_h$, embedding vector size $n_e$, the number of temporal and inter-trajectory attention layers, and the temporal attention score drop rate $p_d$.
As a baseline, we use $(d_h, n_e, p_d) = (16, 4, 0.1)$ and set the number of both temporal and inter-trajectory attention layers to 2. We then vary one hyperparameter at a time and record the highest likelihood achieved during the training process. From the results, we conclude that the baseline choice of $(d_h, n_e, p_d) = (16, 4, 0.1)$ with 2 layers of temporal and inter-trajectory attention yields optimal performance given the current computational resources.

\subsection{Sample complexity study for state distinguishing task}\label{sec:F2}

\begin{figure}[ht]
    \centering
    \includegraphics[page=6, width=\linewidth]{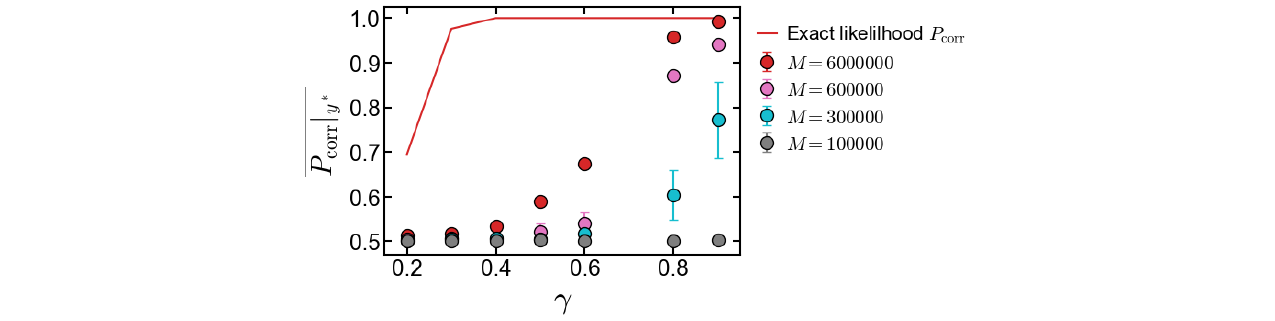}
    \caption{Average likelihood $\overline{P_{\rm corr}|_{y^*}}$ of correctly inferring the initial state as a function of $\gamma$, for different sample complexity $M$. The red line represents the exact likelihood for $N=64$ and system size $L=12$. The error bar represents the standard error over four independently trained models.}
    \label{fig:8}
\end{figure}

In SFig.~\ref{fig:8}, we present a sample complexity study for the state distinguishing task to assess how many measurement trajectories are required for the model to accurately learn the measurement-induced phase transitions (MIPTs).  
We use the QuAN model with the optimal set of hyperparameters and set size $N_{\mathrm{op}} = 64$. The training sample size $M$ is varied across $M = 6,000,000$, $600,000$, $300,000$, and $100,000$, and for each case, we present the sample mean of the average likelihood $\overline{P_{\rm corr}|_{y^*}}$ over 4 independently trained models for different measurement strengths $\gamma$.
To ensure that models trained with smaller $M$ are not disadvantaged due to fewer gradient updates per epoch, we adjust the number of training epochs accordingly: for smaller $M$, we use more epochs so that the total number of optimization steps remains approximately constant across different values of $M$.  
As shown in SFig.~\ref{fig:8}, the likelihood $P_{\rm corr}$ increases with $\gamma$, particularly beyond the critical point $\gamma_c = 0.39$. However, for small training sizes $M < 6,000,000$, the likelihood does not saturate to 1, indicating incomplete learning. In particular, for $M = 100,000$, the model never deviates from the random guess baseline $\overline{P_{\rm corr}|_{y^*}} = 0.5$, suggesting that achieving high classification accuracy near $P_{\rm corr} = 1$ requires a substantially large number of training trajectories.

\subsection{Minimal sample complexity and the system size scaling for phase recognition task}\label{sec:F3}

\begin{figure}[ht]
    \centering
    \includegraphics[page=7, width=0.95\linewidth]{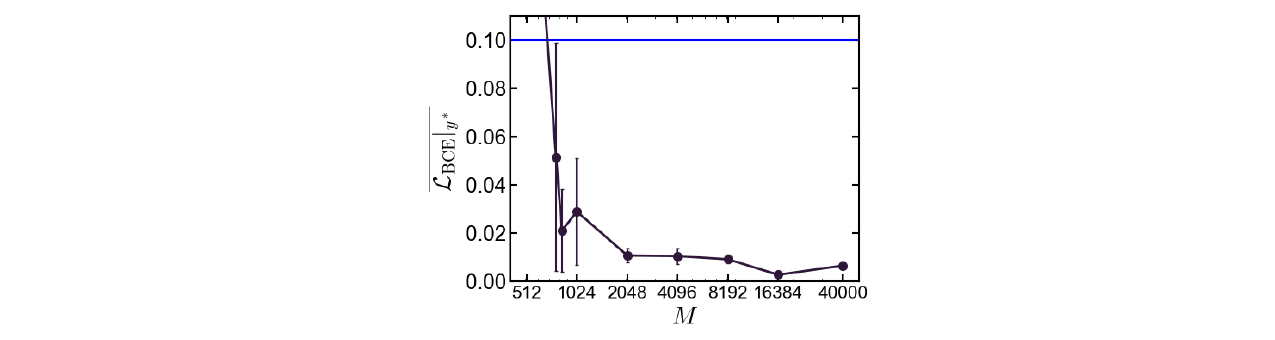}
    \caption{Average test loss as a function of training sample size between $512 \leq M \leq 40000$ for system sizes $L=20$, using noiseless data to train and test the model. The blue horizontal line represents the threshold $\epsilon=0.1$ to determine the minimal sample complexity $M^*$. The error bar represents the standard error over 8 independently trained models.}
    \label{fig:9}
\end{figure} 

\begin{figure}[ht]
    \centering
    \includegraphics[page=8, width=\linewidth]{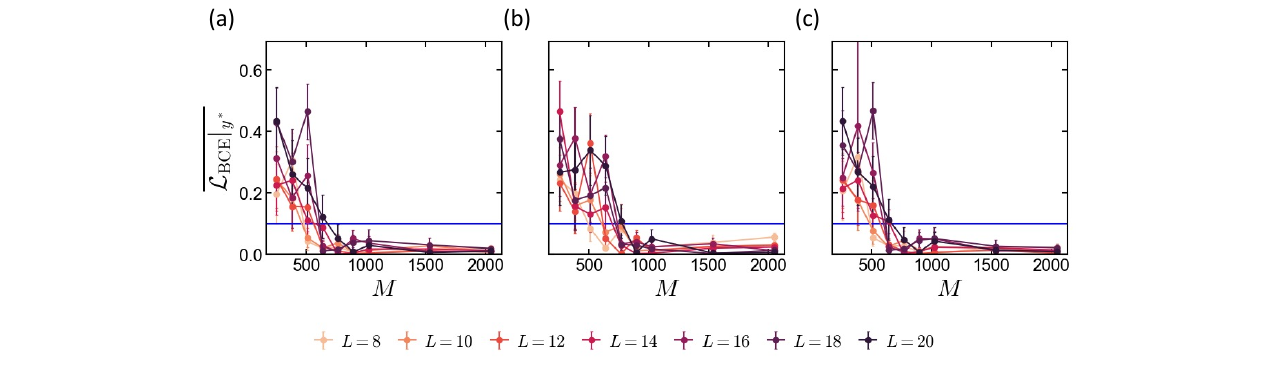}
    \caption{Average test loss as a function of training sample size $M$ for three training-testing modules and different system sizes $L$. (a) Trained on noiseless data and obtained test loss for noiseless data. (b) Trained on noisy data and obtained test loss for noisy data. (c) Trained on noiseless data and obtained test loss for noisy data. Blue horizontal line represents the threshold $\epsilon=0.1$ to determine the minimal sample complexity $M^*$. The error bar represents the standard error over 8 independently trained models.}
    \label{fig:10}
\end{figure}

\begin{figure}[ht]
    \centering
    \includegraphics[page=9, width=\linewidth]{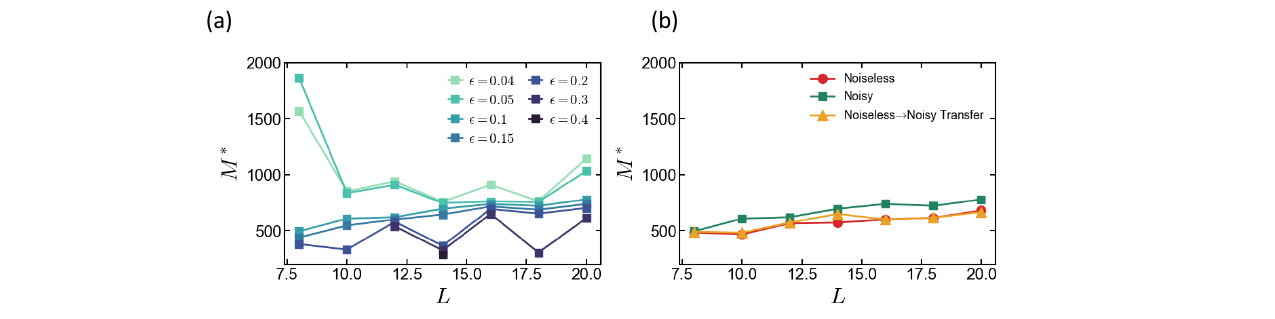}
    \caption{(a) Minimal sample complexity $M^*(\epsilon)$ using different loss threshold $\epsilon$. We use noisy data for training and testing to illustrate our approach. (b) Minimal sample complexity $M^*$ using $\epsilon=0.1$ for three training-testing modules.}
    \label{fig:11}
\end{figure}

In the main text, Fig.~4(a) presents QuAN’s confidence in predicting the strong monitoring phase using $M = 832$ training trajectories. This result is based on the minimal number of trajectories required to witness the measurement-induced phase transition, defined as ``minimal sample complexity''. Here, we provide a detailed analysis of the minimal sample complexity $M^*$ on a broader range of $M$, and justify the choice of $M^*$ in the main text.

To estimate $M^*$, we begin with a coarse scan over a broad range of training sample sizes, from $M = 40,000$ down to $M \sim 512$, to identify the region of interest. This initial scan helps us determine the appropriate range of $M$ values for more detailed analysis. Throughout, we use the same model architecture and optimal hyperparameters used in the state distinguishing task, also fixing the set size at $N_{\mathrm{op}} = 64$.
We use the test loss of the trained model $\mathcal{L}_{\rm BCE}\vert_{y^*}$, calculated over set of trajectories from 4 training $\gamma$ points ($\gamma=0.05, 0.1, 0.85, 0.9$) as a proxy to assess training success, where the loss quantifies how well the predicted label $y^*(\mathbb{m})$ matches the true label $Y$. 
For each $M$, we train the model independently 8 times and compute the average test loss $\overline{\mathcal{L}_{\rm BCE}|_{y^*}(M)}$ using a fixed test set of $M = 10,000$ trajectories. To ensure that models with smaller $M$ are trained with a comparable number of total gradient updates, we increase the number of training epochs accordingly, since smaller $M$ results in fewer updates per epoch.

SFig.~\ref{fig:9} shows the average test loss for noiseless data with $L = 20$. We observe that for $2048 \leq M \leq 40,000$, the test loss remains relatively low, and the variance across independent trainings is small, indicating that the model reliably learns to recognize the weak and strong monitoring phases. Based on this observation, we conduct a more focused scan for $M \leq 2048$ to determine where model training begins to fail.

SFig.~\ref{fig:10}(a–c) presents test loss results in the range $M<2048$ for noiseless, noisy, and transfer settings across different system sizes $L$. Since we use the binary cross-entropy loss, a larger test loss implies training failure. We define the minimal sample complexity as the smallest $M$ for which the average test loss falls below a threshold $\epsilon$:
\begin{align}
M^* = \arg \min_M\left(\overline{\mathcal{L}_{\rm BCE}|_{y^*}(M)} \leq \epsilon\right).
\end{align}
We adopt $\epsilon = 0.1$ as the default threshold for identifying $M^*$.

As shown in SFig.\ref{fig:10}, the test loss transitions from near-zero to high values as we decrease the training sample size $M$. To ensure that our choice of $\epsilon$ captures this transition robustly, we perform a threshold sensitivity analysis in SFig.\ref{fig:11}(a), varying $\epsilon$ from 0.04 to 0.4. Thresholds that are too low render $M^*$ highly sensitive to fluctuations, while overly high thresholds fail to capture the onset of successful learning. Based on this analysis, we conclude that a threshold in the range $0.1 \leq \epsilon < 0.2$ offers a reliable characterization of the minimal sample complexity. We use $\epsilon = 0.1$ in our main analysis, and the resulting $M^*$ values are shown in SFig.~\ref{fig:11}(b) (see also main text Fig.~5(b)).

\subsection{Optimal set size study for phase recognition task}\label{sec:F4}

\begin{figure}[ht]
    \centering
    \includegraphics[page=10, width=0.9\linewidth]{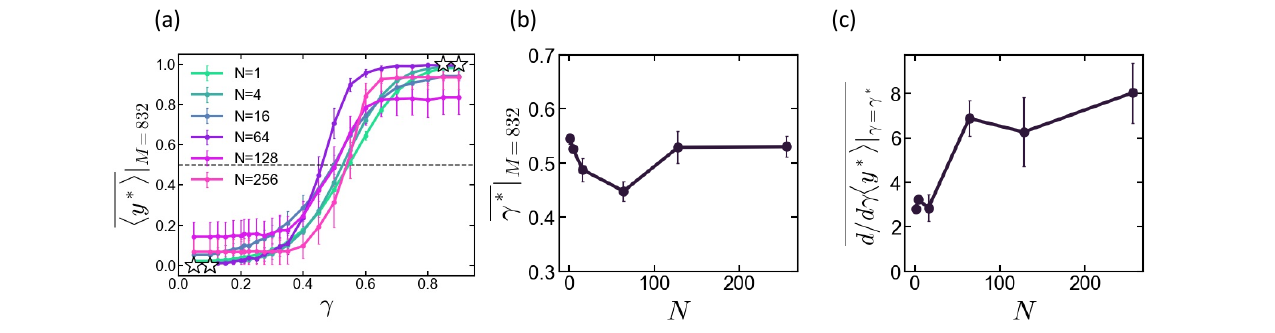}
    \caption{(a) Average strong monitoring phase prediction $\overline{\langle y^*\rangle}$ as a function of weak measurement strength $\gamma$. The error bars represent the standard error over 8 independent model trainings. We use a system size $L=20$ and $M=832$ training samples.  The dashed line marks the point $\langle y^* \rangle=0.5$ where we make the transition prediction $\gamma^*$. (b) Predicted critical point $\overline{\gamma^*}$ as a function of set size $N$, using $M=832$ training samples. (c) First derivative of the phase prediction curve at a predicted critical point as a function of set size $N$. }
    \label{fig:7}
\end{figure}

In SFig.~\ref{fig:7}, we show the effect of the hyperparameter, the set size $N$, in predicting the phase transition point. 
We train the same model used in the state distinguishing study on noiseless data with $M = 832$ trajectories for $L = 20$, varying the set size from $N = 1$ to $N = 256$.
After obtaining the strong monitoring phase prediction curve $\langle y(\gamma) \rangle$ as a function of the measurement strength $\gamma$, we extract the predicted critical point $\gamma^*$ defined by $\langle y(\gamma^*)\rangle = 0.5$, and quantify the sharpness of the transition using the first derivative at $\gamma^*$, denoted by $\overline{d\langle y\rangle/d\gamma |_{\gamma^*}}$. We compute the average predicted critical point and average sharpness across 8 independent trainings.
\begin{align}
    \overline{\gamma^*} &= \frac{1}{8}\sum_r^8 \gamma^*_r \\
    \overline{\frac{d\langle y^*\rangle}{d\gamma}\bigg\vert_{\gamma=\gamma^*}} &= \frac{1}{8}\sum_r^8\frac{d\langle y^*_r\rangle}{d\gamma}\bigg\vert_{\gamma=\gamma^*_r},
\end{align}
where $r$ indexes independently trained models.
From the average strong monitoring phase prediction shown in SFig.\ref{fig:7}(a), we observe that the predicted phase boundary becomes sharper as the set size $N$ increases up to 64. However, for $N \geq 128$, the error bars grow significantly due to the reduced number of available training sets, $\lfloor M/N \rfloor$.
The average predicted critical point $\overline{\gamma^*}$ in SFig.\ref{fig:7}(b) approaches the actual transition point $\gamma_c$ with increasing $N$, but saturates around $N = 64$; set sizes beyond $N > 64$ lead to larger uncertainties in $\overline{\gamma^*}$.
Similarly, in SFig.~\ref{fig:7}(c), the magnitude of the first derivative, or sharpness of the transition, increases with $N$ up to 64, but becomes unreliable for $N > 64$ due to increased variance.
We conclude that the optimal set size for the phase recognition task is $N_{\mathrm{op}} = 64$, consistent with the value identified in the state distinguishing task. At this set size, both the predicted critical point and its sharpness are saturated and stable across independent trainings.

\subsection{System size scaling of the transition point for phase recognition task}\label{sec:F5}

\begin{figure}[ht]
    \centering
    \includegraphics[page=11, width=0.9\linewidth]{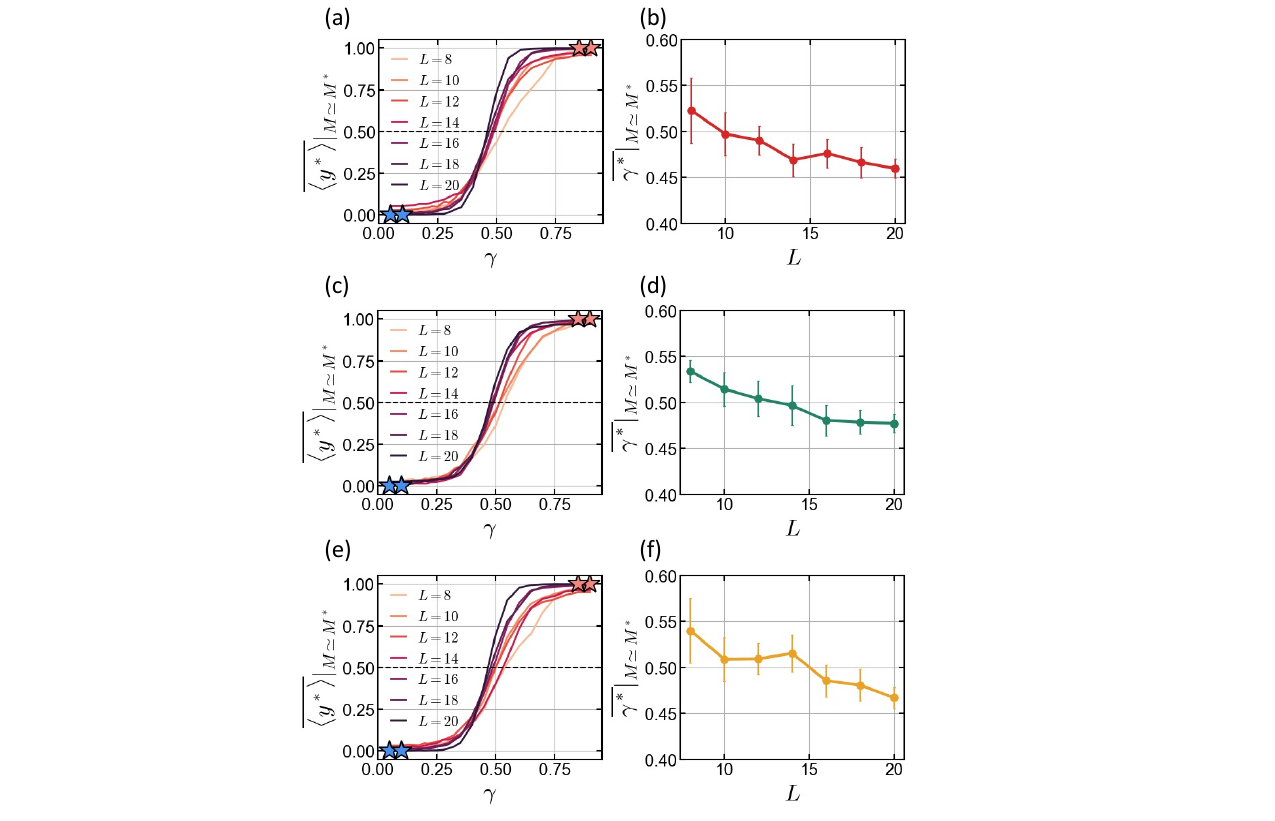}
    \caption{(a,c,e) Average strong monitoring phase prediction $\overline{\langle y^*\rangle}$ as a function of measurement strength $\gamma$ for different system size $L$. We use optimal set size $N_{op}=64$ and minimal sample complexity $M = \lceil M^*(L) / N_{op} \rceil\cdot N_{op}$. (a) Trained and tested on noiseless data. (c) Trained and tested on noisy data. (e) Transferred noiseless data training to noisy data. (b,d,f) Machine-predicted critical point $\overline{\gamma^*}$ as a function of system size $L$ using minimal sample complexity $M=M^*$. The error bar represents the standard error over 8 independent trainings. (b) Trained and tested on noiseless data. (d) Trained and tested on noisy data. (f) Transferred noiseless data training to noisy data. }
    \label{fig:12}
\end{figure}

In SFig.~\ref{fig:12}, we present the system size dependence of the strong monitoring phase prediction curve $\overline{\langle y^*\rangle}$ and the machine-predicted critical point. The model is trained using the minimal sample complexity $M = \lceil M^*(L) / N_{op} \rceil\cdot N_{op}$, and we report the average prediction $\overline{\langle y^*\rangle}$ over 8 independent training runs, and the average critical point $\overline{\gamma^*}$ at $\langle{y(\gamma^*)}\rangle= 0.5$ over 8 independent training runs. 
Consistent with the discussion in Fig.~5(c) of the main text, the predicted critical point decreases as the system size $L$ increases. Additionally, when comparing the average prediction curves across different system sizes, we observe that the prediction becomes sharper with increasing $L$, resulting in a crossing point for varying system sizes.
It is important to emphasize that we use the minimal sample complexity $M \approx M^*(L)$, which explicitly depends on the system size $L$, to make a fair comparison between different system sizes. As previously discussed in Appendix~\ref{sec:F3}, the minimal sample complexity is defined as the smallest number of trajectories required to achieve successful model training. Here, success is specifically defined by achieving an average test loss $\mathcal{L}$ below a certain threshold independent of system size, ensuring reliable identification of the critical point. 
The minimal sample complexity thus reflects model performance, indicating that fewer resources or trajectories are required to train models for smaller system sizes. Therefore, adopting a minimal sample complexity that scales with system size provides a fairer comparison than using a constant sample size across different system sizes.

\section{QuAN accessing the \texorpdfstring{$\Prob(p)$}{Prob(p)} distribution through inter-trajectory attention}\label{sec:G}

\begin{figure}[ht]
    \centering
    \includegraphics[page=13, width=0.9\linewidth]{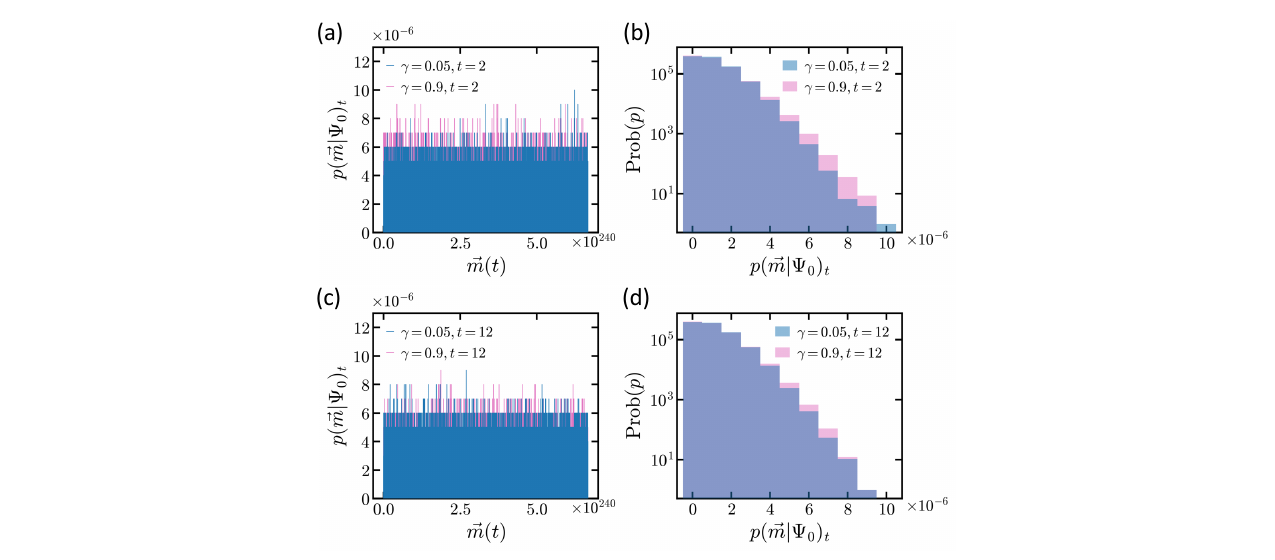}
    \caption{(a) Born probabilities of early-time ($t=2$) measurement outcomes $p(\vec{m}|\Psi_0)_{t=2}$ for $\gamma=0.05$ and $0.9$.
    (b) Distribution of Born probability $\mathrm{Prob}(p)$ for $\gamma=0.05$ and $0.9$, where $p \equiv p(\vec{m}|\Psi_0, \gamma)_{t=2}$ is the probability of a measurement outcome at early-time $t=2$.
    (c) Born probabilities of late-time ($t=12$) measurement outcomes $p(\vec{m}|\Psi_0)_{t=12}$ for $\gamma=0.05$ and $0.9$.
    (d) Distribution of Born probability $\mathrm{Prob}(p)$ for $\gamma=0.05$ and $0.9$, where $p \equiv p(\vec{m}|\Psi_0, \gamma)_{t=12}$ is the probability of a measurement outcome at early-time $t=12$. The long tail present at $t=2$ for $\gamma=0.9$ disappears at $t=12$, indicating the time dependence of the $\Prob(p)$ for the strong monitoring regime. We use $M=50,000$ measurement outcome to estimate Born probabilities. }
    \label{fig:g1}
\end{figure} 

The goal of this section is to address the following question: How does QuAN distinguish between the strong and weak monitoring regimes in the phase recognition task?  
We show that QuAN accesses the tail information of the distribution of the Born probability $\Prob(p)$ through inter-trajectory attention, enabling it to contrast the two phases based on the measurement outcome as input.
In particular, we demonstrate that attention scores increase with the product of Born probabilities at early times, where $\Prob(p)$ exhibits a longer tail in the strong monitoring regime.

We begin by analyzing the underlying statistics of the Born probability $p(m)_t$, estimated as $k_{m(t)}/M$ from empirical frequency (see Appendix~\ref{sec:D}), for strong ($\gamma = 0.9$) and weak ($\gamma = 0.05$) monitoring regimes. 
As shown in SFig.~\ref{fig:g1}, at early time $t=2$, the $\Prob(p)$ distribution for $\gamma = 0.9$ exhibits a longer tail than that for $\gamma = 0.05$, consistent with theoretical expectations from Appendix~\ref{sec:A}. This long tail reflects an increased frequency of large Born probability outcomes in the strong monitoring regime. In contrast, at a late time $t=12$, the two $\Prob(p)$ distributions converge toward a short-tailed Binomial distribution.
These observations suggest that QuAN might leverage the early-time tail structure of $\Prob(p)$ to distinguish between the two phases in the phase-recognition task.

\begin{figure}[ht]
    \centering
    \includegraphics[page=12, width=0.9\linewidth]{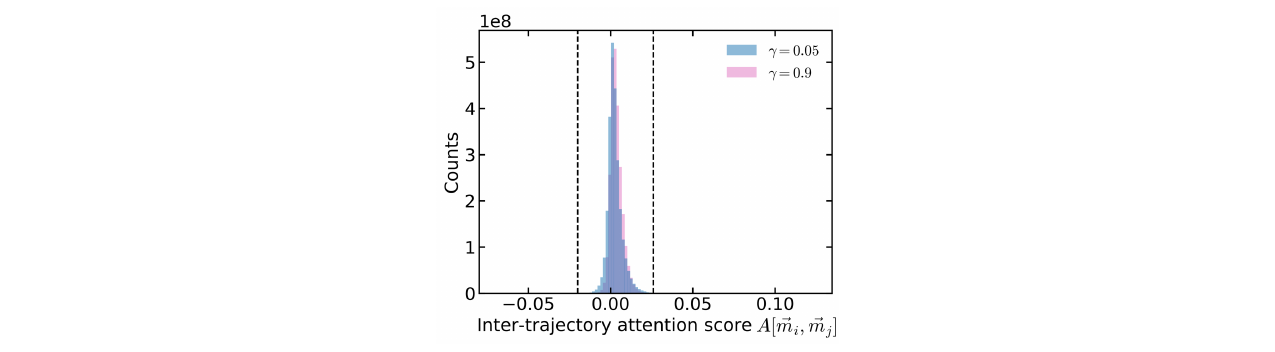}
    \caption{Histogram of inter-trajectory attention scores $A[m_i,m_j]$ for $\gamma=0.05$ (blue) and $\gamma=0.9$ (pink). The two dashed lines indicate the range that contains 99.9\% of the attention scores. Attention scores are computed using all trajectory pairs from the test dataset of size $M=50,000$.}
    \label{fig:g2}
\end{figure} 

Next, we examine how the attention mechanism of QuAN differentiates the two regimes. Specifically, we compute the unnormalized inter-trajectory attention scores,
\begin{align}
    A[m_i,m_j] = \left( Q^{(s)}\texttt{T-Attn}(\texttt{emb}(m_i)\right) \cdot \left( K^{(s)}\texttt{T-Attn}(\texttt{emb}(m_j))\right),
\end{align}
where $\texttt{emb}$ is the embedding layer, $\texttt{T-Attn}$ is the temporal attention block, and $Q^{(s)}, K^{(s)}$ are the query and key matrices of the inter-trajectory self-attention block as defined in Appendix~\ref{sec:E1} (see also Eq.~\eqref{eq:intertrajattentionscore}). 
As shown in SFig.~\ref{fig:g2}, 99.9\% of the inter-trajectory attention scores are distributed between -0.015 and 0.03 out of the whole range (-0.07 and 0.12), and the histogram of inter-trajectory attention scores reveals that the strong monitoring regime ($\gamma=0.9$) exhibits a slight skew toward larger attention values compared to the weak monitoring case ($\gamma=0.05$). 
Although the difference is subtle at the distribution level, this suggests a potential phase-dependent bias in how QuAN assigns inter-trajectory attention scores. (The model and dataset details used for this analysis are summarized in Table~\ref{tab:sab_data_details}.)

\begin{figure}[ht]
    \centering
    \includegraphics[page=14, width=0.9\linewidth]{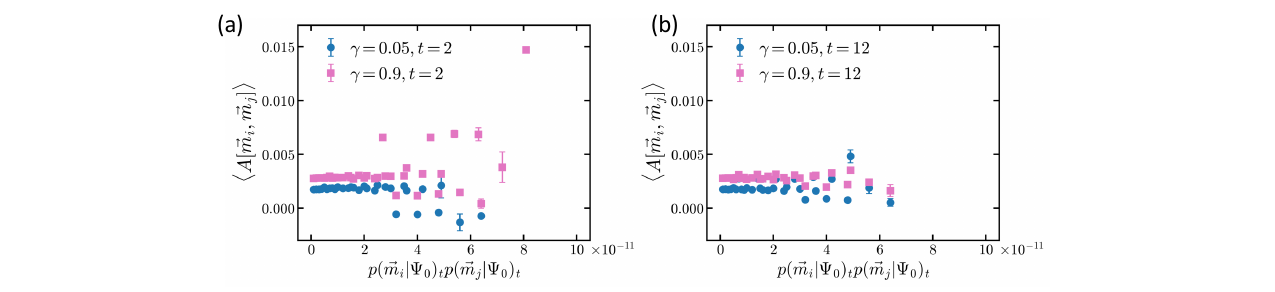}
    \caption{Average inter-trajectory attention score $\langle A[m_i,m_j]\rangle$ as a function of the product of Born probabilities of the pair of measurement outcomes at (a) early time $t=2$ and (b) later time $t=12$. For $\gamma=0.9$ (pink), we observe a clear increasing trend at $t=2$, absent in the weak-monitoring case ($\gamma=0.05$, blue). Attention scores are computed using all trajectory pairs from the test dataset of size $M=50,000$.}
    \label{fig:g3}
\end{figure} 

To further assess whether QuAN's attention selectively focuses on high-probability measurement outcomes -- the aspect of the tail of the $\Prob(p)$ distribution -- we analyze how the inter-trajectory attention score $A[m_i, m_j]$ correlates with the product of Born probabilities $p(m_i)_tp(m_j)_t$. Since the inter-trajectory attention score $A[m_i,m_j]$ involves two trajectories, using the product $p(m_i)_tp(m_j)_t$ allows us to test whether QuAN's attention mechanism correlates with the joint probability of both outcomes, that is, whether QuAN favors to assign high attention scores to pairs of trajectories that are more probable.
In SFig.~\ref{fig:g3}, we plot the average inter-trajectory attention score, which is conditioned under the product of Born probabilities associated with each entry, 
\begin{align}
 \langle A[m_i,m_j]\rangle \equiv \langle A[m_i,m_j]\rangle\Big\vert_{q} = \frac{1}{\sum_{i,j} \delta_{q,\, p(m_i)_tp(m_j)_t}} \sum_{i,j} \delta_{q,\, p(m_i)_tp(m_j)_t} A[m_i,m_j].
\end{align}
As shown in SFig.~\ref{fig:g3}, the average attention score increases with $p(m_i)_tp(m_j)_t$ for $\gamma = 0.9$ at early time $t=2$, where the tail of $\mathrm{Prob}(p)$ is present. This trend diminishes at a later time ($t=12$), and is nearly absent for $\gamma = 0.05$ across both early and late times.
These results indicate that QuAN’s inter-trajectory attention mechanism is sensitive to rare high-probability events in the strong-monitoring regime and early times. This provides direct evidence that QuAN accesses the tail information of $\Prob(p)$ to perform phase recognition.

\begin{table}[ht] \setlength{\tabcolsep}{8pt}
\begin{tabular}{|l|cc|}
\hline
\multicolumn{3}{|l|}{Model training/testing hyperparameter} \\ \hline
Model                       & \multicolumn{2}{|c|}{QuAN (full model)}    \\
Model index out of $8$ trainings     & \multicolumn{2}{|c|}{$7$  } 
 \\
Inter-trajectory attention layer & \multicolumn{2}{|c|}{$1$st layer} \\
System size $L$ & \multicolumn{2}{|c|}{20} \\
Training dataset size             & \multicolumn{2}{|c|}{$M=832$}     \\
Testing dataset size             & \multicolumn{2}{|c|}{$M=50,000$}     \\
\hline
\end{tabular}
\caption{Model hyperparameters and testing data details used in SFig.~\ref{fig:g1} and \ref{fig:g3}.}
\label{tab:sab_data_details}
\end{table}

\bibliography{combined}